\documentclass[12pt,a4paper
]{article}
\usepackage[russian,english]{babel}
\usepackage[cp1251]{inputenc}
\usepackage[T2A]{fontenc}
\usepackage{latexsym}
\usepackage{amssymb}
\usepackage{amsmath}
\usepackage{graphicx}

\newcommand{\be}{\begin{equation}}
\newcommand{\ee}{\end{equation}}
\newcommand{\p}[1]{(\ref{#1})}
\def\a{\alpha}

\def\b{\beta}
\def\p{\partial}

\def\t{\theta}
\def\tr{{\rm tr}}
\def\tr{{\rm tr}\,}
\def\Tr{{\rm Tr}\,}
\def\cN{{\cal N}}
\def\cD{{\cal D}}
\def\cA{{\cal A}}

\def\bea{\begin{eqnarray}}
\def\eea{\end{eqnarray}}

\def\cN{{\cal N}}

\def\cF{{\cal F}}
\def\cH{{\cal H}}
\def\f{\frac}

\def\tr{{\rm tr}\,}

\def\s{\sigma}

\def\d{\delta}
\def\q{\quad}
\def\g{\gamma}

\def\l{\label}
\def\ve{\varepsilon}

\def\cL{\cal L}
\def\O{\Omega}
\def\sB{\stackrel{\frown}{\Box}}
\def\r{\rho}
\def\l{\ldots}
\def\cF{\cal F}
\def\cH{\cal H}
\def\cV{\cal V}
\def\cW{\cal W}
\def\cY{\cal Y}

\date{\it  }
\begin{document}

\begin{center}
\vspace{1cm} {\Large\bf Construction of 6D supersymmetric field
models in $\cN=(1,0)$ harmonic superspace

\vspace{1.2cm} }

\vspace{.2cm}
 {I.L. Buchbinder$^{a}$,
 N.G. Pletnev$^{b}$
}

\vskip 0.6cm { \em \vskip 0.08cm \vskip 0.08cm $^{a}$Department of
Theoretical Physics, Tomsk State Pedagogical University,\\ Tomsk,
634061 Russia; joseph@tspu.edu.ru\\
and\\ National Research Tomsk State University, Tomsk, 634050 Russia
\vskip 0.08cm \vskip 0.08cm
$^{b}$Department of Theoretical Physics, Sobolev Institute of Mathematics\\
and National Research Novosibirsk State University,\\
 Novosibirsk, 630090 Russia; pletnev@math.nsc.ru

 }
\vspace{.2cm}
\end{center}

\begin{abstract}
We consider the six dimensional hypermultiplet, vector and tensor
multiplet models in (1,0) harmonic superspace and discuss the
corresponding superfield actions. The actions for free (2,0) tensor
multiplet and for interacting vector/tensor multiplet system are
constructed. Using the superfield formulation of the hypermultiplet
coupled to the vector/tensor system we develop an approach to
calculation of the one-loop superfield effective action and find its
divergent structure.

\end{abstract}

\section{Introduction}

The construction of the non-Abelian $(1,0)$ and $(2,0)$
superconformal theories in $6D$ has attracted much attention for a
long time (see e.g. \cite{20}, \cite{LP}). Such models are
considered as the candidates for dual gauge theories of the
interacting multiple M5-branes \cite{MMM} and can be related to
near-horizon AdS$_7$ geometries. A crucial ingredient of this
construction is the non-Abelian tensor multiplet gauge fields
\cite{BHS}\footnote{Recently, a number of papers devoted to
constructing the class of non-Abelian superconformal (1,0) and (2,0)
theories in six dimensions have appeared (see \cite{Lambert} and
references therein). These works were inspired by papers \cite{LP},
which explored the 3-algebra gauge structure and used a
non-propagating vector field of negative scaling dimension which
transforms nontrivially under the non-Abelian gauge symmetry.}. A
solution to this problem has been found \cite{SSWW} in the framework
of a tensor hierarchy \cite{TH} which, besides the Yang-Mills gauge
field and the two-form gauge potentials of the tensor multiplet,
contains the non-propagating three- and four-forms gauge potentials.
Construction of the (2,0) models can be realized on the base of
coupling the (1,0) non-Abelian tensor/vector models to the
superconformal hypermultiplets \cite{SSW}. We also mention the work
\cite{Akyol} where the Killing spinor equations of 6-dimensional
(1,0) superconformal theories have been solved and the solutions for
the configuration of the background fields preserving 1,2, 4 and 8
supersymmetries have been found.

Superfield formulation of the tensor hierarchy has been studied in
the paper \cite{IB} where a set of constraints on the
super-$(p+1)$-form field strengths of non-Abelian super-$p$-form
potentials in (1,0) $D6$ superspace has been proposed. These
constraints restrict the field content of the super-$p$-forms to the
fields of the non-Abelian tensor hierarchy. The superfield
formulation of the tensor hierarchy sheds light on a supersymmetric
structure of the theory and can serve as a base for the various
generalizations. They can be useful for searching the superfield
action \cite{action} and for  studying the (2,0) superconformal
theory by superspace methods. However, the superfield Lagrangian
formulation of the theory under consideration has not been
constructed so far.

In this paper we are going to develop the superfield methods for
studying the open problems related to superfield formulation of the
vector/tensor system and calculating the quantum effective action.
Our consideration is based on the harmonic superspace technique
formulated for four dimensions in \cite{gikos}, \cite{gios} and
extended to six dimensions in \cite{Z86}, \cite{HSW}. The superfield
realization of the unitary representations of $(n,0)$ superconformal
algebras $OSp(8{}^\star/2n)$ in six dimensions \cite{GT} has been
found in \cite{FS} and it was shown that the $D6$, (1,0) and (2, 0)
tensor multiplets are described by the analytic superfields in
appropriately defined harmonic superspaces \cite{PH}. In this paper
we are going to demonstrate that a harmonic superspace formalism can
be efficiently implemented for the superfield Lagrangian
construction of the tensor hierarchy models.

The paper is organized as follows. Section 2 is devoted to the basic
notations of the $6D$ harmonic superspace. In section 3 we are going
to review the superfield formulations of the $6D$ (1,0)
hypermultiplet \cite{HSW}, the vector multiplet \cite{Z86} and the
tensor multiplet \cite{sok}, in harmonic superspace. We also discuss
the structure of the (2,0) tensor multiplet. The material of section
3 is used in the other sections to formulate the new superfield
models. Section 4 is devoted to the superfield Lagrangian
construction of the (2,0) tensor multiplet in terms of the (1,0)
hypermultiplet and the (1,0) tensor multiplet. In section 5 we are
going to study the superfield Lagrangian formulation for the
non-Abelian vector/tensor system. We begin with a harmonic
superspace reformulation of the results of the paper \cite{IB}, then
we propose the superfield action for the superconformal models of
tensor hierarchy and, using the results of the section 5, we derive
the component structure of the superfield action and show that it
coincides with the component Lagrangian which was constructed in
\cite{SSWW}. Section 6 will demonstrate a power of superfield
methods. This section is devoted to a study of the quantum effective
action in the (1,0) hypermultiplet theory coupled to the Abelian
vector/tensor system. We are going to develop the superfield proper
time technique, which will allow us to calculate the effective
action in manifestly supersymmetric and gauge invariant form, and
calculate the divergent part of the effective action. We will prove
that this divergent part contains a term, providing the charge
renormalization in the vector/tensor action from Section 5, and a
higher derivative action, found in \cite{ISZ}. In conclusion we will
summarize the results obtained. In Appendix A we will describe the
basic notations and conventions of $6D$ supersymmetry. Appendix B
contains some details of deriving the component action from
superfield action of vector/tensor system.

\section{$6D$, $\cN=(1,0)$ harmonic superspace}

\def\theequation{2.\arabic{equation}}
\setcounter{equation}{0}

Harmonic superspace is a powerful formalism for the off-shell
construction of extended supersymmetric field theories in four and
six dimensions \cite{gikos}, \cite{gios}, \cite{Z86}, \cite{HSW}. In
this section we will briefly describe the basic notations and
conventions which are used in this paper (see the details of $D=6$
superspace e.g. in \cite{HST})\footnote{Harmonic superspace is
closely related to projective superspace which is also successfully
applied to off-shell formulations of extended supersymmetric
theories (see e.g the recent papers \cite{rocek}, \cite{SKD5},
\cite{project}).}.

It is well known that in six dimensions there are two independent
supersymmetry generators. Therefore, the representations of the $6D$
superalgebra are defined by two integers $(p,q)$ \cite{HST}. The
corresponding supersymmetries are generally denoted as ${\cal
N}=(p,q)$ or simply $(p,q)$. In this paper we will construct the
harmonic superfield models corresponding to ${\cal N}=(1,0)$ and
${\cal N}=(2,0)$ supersymmetries.

The six-dimensional superspace is parameterized by the coordinates
$z=\{x^{\a\b}\equiv x^{m}\g^{\a\b}_m, \t^\a_I\}$. Here the odd
coordinates $\t^\a_I$ ($\a$ = 1,\l,4) are the right-handed chiral
spinors of the group $SU^\ast(4)\sim SO(1,5)$ (left-handed spinors
are denoted as $\psi_{\a}$). The index $I$ is the spinor one of the
group $USp(2n)$ (below we use $n=1,2$) and $n$ corresponds to the
$\cN=(n,0)$ supersymmetry. The properties of the matrices
$\g^{\a\b}_m$ are given in the Appendix A. The index $I$ is raised
and lowered with the help of the $USp(2n)$ matrix $\O^{IJ}$ (see the
properties of this matrix e.g. in \cite{BP}),
$\psi_I=\O_{IJ}\psi^J,\q \psi^I=\O^{IJ}\psi_J,\q
\O^{IJ}\O_{JK}=\d^I_K$. The Grassmann coordinates obey the reality
condition $\overline{\t^\a_I}=\t^{\a I}=\O^{IJ}\t^\a_J.$ The basic
spinor derivatives of the $6D, \cN=(n,0)$ superspace are
\be\label{salg}
D_\a^I=\f{\p}{\p\t^\a_I}-i\t^{I\b}\p_{\a\b},\q\{D_\a^I,D^J_\b\}=-2i\O^{IJ}\g^m_{\a\b}\p_m.\ee
The symmetry group of the superspace involves  $USp(2n)$
transformations of the R-symmetry.

The harmonic  $D6$, $\cN=(1,0)$ superspace was introduced in
\cite{Z86}, \cite{HSW},  \cite{ISZ} and it is parameterized by the
coordinates $(x^m, \t^{\a i}, u^{\pm i})$, where harmonics $u^{\pm
i}$ $(\widetilde{u^\pm_i}=u^{\pm i}, u^{+ i}u^-_i=1 (i=1,2))$ live
on the coset R-symmetry of the group $SU(2)/U(1).$ Besides the
standard (or central) basis ($x^m,\t^\a_i, u^\pm_i$) one can
introduce the analytical basis ($\zeta^M_A=\{x^m_A, \t^{+ \a}\},
u^\pm_i, \t^{-\a} $): \be x^a_A=x^a +i\t^-\g^a\t^+,\q \t^{\pm
\a}=u^\pm_i\t^{\a i}.\ee The important property of the coordinates
$\zeta^M_A, u^\pm_i$ is that they form a subspace closed under
$\cN=(1,0)$ supersymmetry transformations. The covariant harmonic
derivatives which form  the Lie algebra of $SU(2)$ group ($ [D^{++},
D^{--}]=D^0$) in the analytic basis have the form
$$ D^{++}=u^{+i}\f{\p}{\p u^{-i}}+i\t^+\not\!\p\t^+ +\t^{+\a}\f{\p}{\p\t^{-\a}}, \q D^{--}=u^{-i}\f{\p}{\p u^{+i}}+i\t^-\not\!\p\t^- +\t^{-\a}\f{\p}{\p\t^{+\a}}~,$$
$$D^0=u^{+i}\f{\p}{\p u^{+i}}-u^{-i}\f{\p}{\p u^{-i}}+\t^{+\a}\f{\p}{\p\t^{+\a}}-\t^{-\a}\f{\p}{\p\t^{-\a}}.$$
by using the analytic subspace, one can define the analytical
superfields, which do not depend on $\t^{-\a}$, i.e. they satisfy
the condition of the Grassmann analyticity $D^+_\a \phi=0,$ where
the spinor derivatives in the analytic basis have the form \be
D^+_\a=\f{\p}{\p\t^{-\a}},\q
D^-_\a=-\f{\p}{\p\t^{+\a}}-2i\p_{\a\b}\t^{-\b}, \q
\{D^+_\a,D^-_\b\}=2i\p_{\a\b}.\ee The possibility to formulate the
theory in terms of G-analytic superfields is a crucial advantage of
the harmonic superspace formalism \footnote{ In some cases it may be
helpful to use an anti-analytic basis in which
$$x^a_{\bar{A}}=x^a-i\t^+\g^a\t^-,\q D^-_\a=-\f{\p}{\p\t^{+\a}}, \q D^+_\a=\f{\p}{\p\t^{-\a}}-2i\p_{\a\b}\t^{+\b}$$
$$D^{++}=u^{+i}\f{\p}{\p u^{-i}}+\t^{+\a}\f{\p}{\p\t^{-\a}}-i\t^+\p\t^+,\q D^{--}=u^{-i}\f{\p}{\p
u^{+i}}+\t^{-\a}\f{\p}{\p\t^{+\a}}-i\t^-\p\t^-$$}.

\section{Harmonic superfields and their interactions}

\def\theequation{3.\arabic{equation}}
\setcounter{equation}{0}

It is known that the massless conformal (1,0) and (2,0) superfields
in six dimensions are divided into two classes: (i) the superfields
whose first component carries any spin but it is an $USp(2n)$
singlet; (ii) the 'ultrashort' analytic superfields in harmonic
superspace, their first component is a Lorentz scalar but it carries
$USp(2n)$ indices \cite{FS}. All these superfields satisfy some
superspace constraints. In this paper we will consider the simplest
superfields from both of the above classes, which correspond to the
following three types of (1,0) 6D multiplets: the hypermultiplet,
the vector multiplet and the tensor multiplet.

\subsection{Hypermultiplet}
The $(1,0)$ and $(2,0)$ hypermultiplets are described by the
superfields $q^I(x,\t)$ and their conjugate $\bar{q}_I(x,\t), \q
\bar{q}_I=(q^I)^\dag$, both in the fundamental representation of
$USp(2n)$ group. Here $i=1,2$ for the $(1,0)$ case and $I=1,...,4$
for the $(2,0)$ case. The corresponding constraint is
\be\label{10q}D_\a^{(I}q^{J)}(x,\t)=0~.\ee

In the case of $\cN=(1,0)$ supersymmetry, the superfield $q^i(x,\t)$
has a short expansion $q^i(z)=f^i(x)+\t^{\a i}\psi_\a(x) +\ldots.$
The doublet of scalars $f^i$ and the spinor $\psi_\a$ satisfy the
equations $\Box f^i=0, \q \p^{\a\b}\psi_\b=0.$ As a result, the
$\cN=(1,0)$ hypermultiplet in six dimensions has 4 bosonic+4
fermuinic real degrees of freedom.

The off-shell Lagrangian formulation of the  hypermultiplet is based
on the use of the analytic superfields in harmonic superspace. In
this formulation the hypermultiplet is described by an unconstrained
analytic superfield $q_A^+(\zeta, u)$ satisfying the reality
condition $\widetilde{(q^{+A})}\equiv q^+_A=\ve_{AB}q^{+B}$ \be
D^+_\a q_A^+(\zeta,u)=0~.\ee Here $A=1,2$ is a Pauli-G\"ursey index
lowered and raised by $\ve_{AB},\ \ve^{AB}.$ After the expansion of
$q^+$ in $\t^+$ and $u$ we obtain an infinite set of auxiliary
fields which vanish on-shell due to the equations of motion \be
D^{++}q^{+}(\zeta, u)=0~.\ee The equations of motion follows from
the action \be\label{actHyper} S_q=-\f12\int d\zeta^{(-4)}du \
{q}^{+A} D^{++}q^{+}_A~.\ee Here
$d\zeta^{(-4)}=d^6xd^{4}\theta^{+}.$ This formulation allows us to
write down the most general self-couplings in the form of the
arbitrary potential ${\cal L}^{(+4)}(q^+, \tilde{q}^+)$ \cite{gios}.
The corresponding sigma models have the complex hyper-K\"ahler
manifolds as their target manifolds \cite{hyper}.

\subsection{Vector multiplet}
The off-shell (1,0) non-Abelian vector supermultiplet is realized in
$6D$ conventional superspace as follows \footnote{An incomplete list
of the references includes \cite{Z86}, \cite{HSW}, \cite{ISZ},
\cite{HST}, \cite{z}, \cite{zupnikZero}.}. As usual one introduces
the gauge-covariant derivatives
$$\cD_M=D_M+\cA_M ,\q [\cD_M,\cD_N]=T_{MN}^{\ \ L}\cD_L+F_{MN},$$
with $D_M = \{D_m,D^i_\a\}$ being the flat covariant derivatives
obeying the anti-commutation relations (\ref{salg}),  and $\cA_M$
being the gauge connection taking the values in the Lie algebra of
the gauge group. The gauge-covariant derivatives under consideration
obey the constraints  $F_{\a\b}^{ij}=0$ and \be
\{\cD_\a^i,\cD^j_\b\}=-2i\ve^{ij}\cD_{\a\b},\q
[\cD_\g^i,\cD_{\a\b}]=-2i\ve_{\a\b\g\d}W^{i\d}.\ee Here $W^{i\a}$ is
the superfield strength  of the  anti-Hermitian superfield gauge
potential obeying the Bianchi identities.

The constraints are solved in the framework of the harmonic
superspace. In this case, the integrability condition
$\{\cD^+_\a,\cD^+_\b\}=0$ yields $\cD^+_\a=e^{-ib}D^+_\a e^{ib}$
with some Lie-algebra valued harmonic superfield $b(z,u)$ of zero
harmonic U(1) charge. In the $\lambda$-frame, the spinor covariant
derivatives $\cD^+_\a$ coincide with the flat ones,
$\cD^+_\a=D^+_\a=\f{\p}{\p\t^{-\a}}$, while the harmonic covariant
derivatives acquire the connection $V^{++}$,
\be\label{cD++}\cD^{++}=D^{++}+V^{++}~.\ee The connection is an
unconstrained analytic potential of the theory. In the Wess-Zumino
gauge, the component expansion of $V^{++}(\zeta, u)$ \be
V^{++}_{WZ}=\t^{+\a}\t^{+\b}A_{\a\b}(x_A)+(\t^+)^3_\a\lambda^{-\a}(x_A)+3(\t^+)^4Y^{--}(x_A)~,\ee
involves  the  physical gauge fields and the auxiliary fields.

The other non-analytic harmonic connection $V^{--}(z, u)$ is
uniquely determined in terms of $V^{++}$ as a solution of the
zero-curvature condition \cite{z}, \cite{gios}
\be\label{zeroc}D^{++}V^{--}-D^{--}V^{++}+[V^{++},V^{--}]=0.\ee The
connection $V^{--}$ transforms as $\d V^{--}=-\cD^{--}\Lambda$ under
gauge transformations. Here the gauge parameter $\Lambda$ is an
analytic anti-Hermitian superfield.

Using the connection $V^{--}$ one can build the spinor and the
vector superfield connections as follows
$${\cal A}^-_\a=-D^+_\a V^{--},\q {\cal A}_{\a\b}=\f{i}{2}D^+_\a D^+_\b V^{--}~.$$
This yields \be\label{defW}W_\lambda^{+\a}=-\f14 (D^+)^{3\a}
V^{--},\ee where $W^{\a i}_\lambda$ is the field strength in the
$\lambda$-frame. The Bianchi identities lead to relations \be D^+_\a
W^{-\a}=D^-_\a W^{+\a}, \q D_\a^\pm
F_{ab}=iD_{[a}(\g_{b]})_{\a\b}W^{\pm\b}.\ee The vector superfield
strength is defined as follows $ F_\a^\b=(D^-_\a W^{+\b}-D^+_\a
W^{-\b}).$ The other useful consequences of the Bianchi identities
are
$$ D^+_\b
W^{+\a}=\f14\d^\a_\b Y^{++},\q Y^{++}=-(D^+)^4V^{--},\q \cD^{++}
Y^{++}=0, $$
$$\q W^-_\a=\cD^{--}W^+_\a, \q \f12\cD^{--}Y^{++}=D^-_\a W^{+\a}.$$
\be\label{identity} D^{++}W^{+\alpha}=0. \ee These relations define
the superfield $Y^{++}$ which will be used further.

The superfield action of $6D$ SYM theory is written in the form
\be\label{actZ}S_{SYM}=\f{1}{f^2}\sum_{n=1}^\infty\f{(-1)^{n+1}}{n}\tr\int
d^{14}z du_1\ldots du_n\f{V^{++}(z,u_1)\ldots
V^{++}(z,u_n)}{(u^+_1u^+_2) \ldots (u^+_{n}u^+_1)}.\ee Here $f$ is
the dimensional coupling constant ($[f]=-1$). The corresponding
equations of motion have the form $Y^{++}=(D^+)^4V^{--}=0.$ The
component fields of $W^{+\a}$ and $V^{++}$ are related to each other
with help of the zero-curvature condition (\ref{zeroc}) and due to
the definition (\ref{defW}).

It is known that the superfield action with dimensionless coupling
constant \cite{ISZ} has the form \be\label{SISZ} S =
\f{1}{2g^2}\tr\int d\zeta^{(-4)}du \ (Y^{++})^2. \ee It possesses
the superconformal symmetry and contains the higher derivatives.

To conclude this section, we give the decomposition of the
superfield $V^{--}$ in terms of the component fields
\be\label{potSYM}V^{--}(x_A,\t^-,\t^+,u)=\t^{-\a}\t^{-\b}v_{\a\b}(x_A,
\t^+)+(\t^-)^3_\a v^{+\a}(x_A, \t^+)+(\t^-)^4v^{++}(x_A, \t^+),\ee
$$ v_{\a\b}=A_{\a\b}+\f14\ve_{\a\b\g\d}\t^{+\g}\lambda^{-\d}-\f{1}{4}\ve_{\a\b\g\d}\t^{+\g}\t^{+\d}Y^{--},$$
$$v^{+\a}=-\f12\lambda^{+\a}+\t^{+\a}Y^{+-}+\t^{+\b}if_\b^\a +\t^{+\g}\t^{+\d}\d^\a_{[\g}\omega^-_{\d]}+(\t^+)^3_\b\kappa^{(-2)\b\a}+(\t^+)^4\s^{(-3)\a},$$
$$ \f12f_\b^\a=(\g_{mn})^\a_\b F_{mn}, \q F_{mn}=\p_mA_n-\p_nA_m-i[A_m,A_n],\q \cD_{\a\b}=\p_{\a\b}-i[A_{\a\b}, \cdot],$$$$ 3\omega^-_\a=i\cD_{\a\b}\lambda^{-\b},\q\q
\f12\ve_{\a\b\g\d}\kappa^{(-2)\g\d}=2i\cD_{\a\b}Y^{--}
+\f{1}{4}\ve_{\a\b\g\d}\{\lambda^{-\g},\lambda^{-\d}\},$$
$$v^{++}=Y^{++}+\t^{+\a}\chi^{+}_\a+\t^{+\a}\t^{+\b}\O_{\a\b}+(\t^+)^3_\a\r^{-\a}+(\t^+)^4\pi^{(-2)},$$
$$\chi^+_\a=i\cD_{\a\b}\lambda^{+\b},\q \ve^{\a\b\g\d}\O_{\g\d}=-2i\cD^{\a\b}Y^{+-}-\cD^{[\a\g}f^{\b]}_\g-\f{1}{4}\{\lambda^{+[\a},\lambda^{-\b]}\},\q
\cD^{(\a\d}f_\d^{\b)}=0,
$$
$$\r^{-\a}=2i\cD^{\a\b}\chi^-_\b+\f{1}{2}[\lambda^{+\a},Y^{--}]-[\lambda^{-\a},Y^{+-}]+i[\lambda^{-\b},f_\b^\a],
$$
$$\ \pi^{(-2)}=\cD^{\a\b}\cD_{\a\b}Y^{--}-\f12\{\lambda^{-\a},\cD_{\a\b}\lambda^{-\b}\}+3[Y^{+-},Y^{--}]~.$$
The relation (\ref{potSYM}) defines the complete component structure
of the superfield $V^{--}$ in terms of the components of the
superfield potential $V^{++}$. The component at
$(\theta^{-})^4(\theta^{+})^4$ has already been calculated in
\cite{ISZ} (at the different conventions). Further we will need all
the components of $V^{--}(x_A,\t^+,\t^-, u)$.

\subsection{Linear multiplet in harmonic superspace}

In this subsection we will briefly review a self-dual tensor
multiple and its description in harmonic superspace \cite{sok}.

As it is well known, the so-called self-dual tensor multiplet
contains a scalar $\phi$, a spinor $\psi_{i\a}$ and an antisymmetric
tensor $B_{ab}$ subject to the self-dual constraint
\be\p_{[a}B_{bc]}-\f16\ve_{abcdef}\p^{d}B^{ef}=0.\ee There are two
ways to describe the self-dual tensor multiplet in harmonic
superspace.

Firstly, one introduces the superfield $\Phi(x,\t)$ subject to the
constraint \be\label{constPhi}D^{(i}_\a D^{j)}_\b\Phi=0~.\ee Such a
superfield is also called a linear. This superfield has no external
indices and obey the reality condition $\overline{\Phi}=\Phi$. In
the case of $\cN=(1,0)$ supersymmetry, the component expansion of
the superfield $\Phi(x,\t)$ has the form
\be\Phi=\phi+\t^\a_i\psi^i_\a+\t^{\a i}\t^\b_iG_{(\a\b)}+\ldots~,\ee
where the component fields satisfy the massless equations of motion.
Note that the field $G_{(\a\b)}$ is related to 3-form field
$G_{abc}$, $G_{(\a\b)}(x)=(\g^{abc})_{\a\b}G_{abc}(x)$ and is
self-dual by definition. We see that on shell, the linear superfield
contains all components of self-dual tensor multiplet.

The above (1,0) self-dual tensor multiplet (with the constraint
(\ref{constPhi})) is formulated in harmonic superspace \cite{sok}
with the help of real superfield, which satisfy the constraints
\be\label{kin} D^+_\a D^+_\b \Phi=0, \q D^{++}\Phi=0.\ee The first
of them means that $\Phi(x,\t^+,\t^-,u)$ is linear in $\t^-$:
\be\label{Phi}\Phi= l(x_A,\t^+,u)+\t^{-\a}f^+_{\a}(x_A,\t^+,u), \ee
where the coefficient functions are the analytic superfields \be
l=\phi+\t^+\psi^-, \q f^+_\a =-\psi^+_\a
-\t^{+\b}i\p_{\a\b}\phi+\t^{+\b}G_{(\a\b)}-i\p_{\a\b}\t^{+\b}\t^{+\g}\psi^-_\g
~. \ee

The dynamical equations follow from the second constraint
(\ref{kin}) which reduces the harmonic dependence of $l$ and $f^{+}$
to a polynomial and thus produces a finite supermultiplet. It leads
to the following harmonic constraints
\be\label{EqPhi}\hat{D}^{++}l+\t^{+\a} f^+_\a=0, \q \q
\hat{D}^{++}f^+_\a=0,\ee from which we obtain the equations of
motion for the components of the self-dual tensor multiplet
$${\p}^{\a\b}\psi^+_\b=0, \q \Box\phi=0, \q {\p}^{\a\g}G_{\g\b}=0~.$$
Note that all these components are the field strengths. 
Besides, the kinematical constraints (\ref{kin}) is solved
\cite{sok} by introducing the prepotential
\be\Phi=(D^+)^{3\a}\Phi^{(-3)}_\a.\ee

Another way to formulate the tensor multiplet in superfield form is
based on the superfield ${\cV}^{\a i}$ subject to the kinematical
constraints \cite{sok} \be\label{cV} D^{(i}_\b
{\cV}^{j)\a}-\f14\d^\a_\b D^{(i}_\g {\cV}^{j)\g}=0.\ee In the
harmonic superspace one gets the superfield ${\cV}^{+\a}(x,
\t^+,\t^-,u)$ under the following constraints \be\label{kinV}
D^+_\a{\cV}^{+\b}-\f14\d^\b_\a D^{+}_\g {\cV}^{+\g}=0, \q
D^{++}{\cV}^{+\a}=0~.\ee In the analytic basis we have
\be\label{potTensor}{\cV}^{+\a}(x_{{A}},\t^-,\t^+,u)=v^{+\a}(\t^+)+\t^{-\a}v^{(+2)}(\t^+)~,\ee
here $v^{+\a}$ and $v^{(+2)}$ are the arbitrary analytic
superfields. The second dynamical constraint $ D^{++}{\cV}^{+\a}=0$
then becomes \be\label{EqB}\hat{D}^{++}v^{(+2)}=0, \q\q
\hat{D}^{++}v^{+\a}+\t^{+\a}v^{(+2)}=0~.\ee The component expansions
of these superfields are obtained from the above relations in the
form
$$v^{(+2)}=f^{(+2)}+\t^{+\a}\kappa^+_\a+(\t^+)^{2\a\b}a_{\a\b}+(\t^+)^3_\a\tau^{-\a}+(\t^+)^4C^{(-2)}~,$$
$$v^{+\a}=\rho^{+\a}+\t^{+\b}(B^\a_\b+\d^\a_\b\Sigma)+
(\t^+)^{2\b\g}\omega^{-\a}_{\b\g}+(\t^+)^3_\b
E^{(-2)\b\a}+(\t^+)^4\varphi^{(-3)\a}~.$$ The kinematical
constraints  (\ref{kinV}) are solved by introducing the prepotential
\cite{sok}: \be {\cV}^{+\a}=(D^+)^{3\a}{\cV}^{(-2)}~,\ee where
\be\label{Vtensor}{\cV}^{(-2)}=(\t^-)^3_\a
v^{+\a}+(\t^-)^4v^{(+2)}~.\ee This prepotential is defined up to the
Abelian gauge transformations \be\d_\Lambda {\cV}^{(-2)}=
-D^{--}\Lambda.\ee If
$$\Lambda\sim \l+
\f12\t^{-\a}\t^{+\b}i\Lambda_{\a\b}+\ve_{\a\b\g\d}\t^{-\a}\t^{-\b}\t^{+\g}\r^{+\d}+(\t^-)^3_\a\t^{+\a}f^{++}+\l~,$$
then
$$
\d {\cV}^{+\a} \sim
\t^{+\b}\p^{\a\g}\Lambda_{\g\b}-\r^{+\a}-\t^{-\a}f^{+2}, \q \d
B_\b^\a=\p^{\a\g}\Lambda_{\g\b}
-\f{1}{4}\d_\b^\a\p^{\g\d}\Lambda_{\d\g}~. $$ The fields
$f^{ij}(x),$ $\rho^\a_i$ form a multiplet of gauge degrees of
freedom, they can be excluded by an appropriate gauge choice, i.e.
one can use the Wess-Zumino gauge.

By substituting the quantities $v^{+\a}, v^{(+2)}$ in (\ref{EqB}) we
find the general solution
$$\tau^{i\a}= 2i\p^{\a\b}\kappa^i_\b,\q \omega^{-\a}_{\b\g}=
-\f12\d^\a_{[\b}\kappa^-_{\g]}, \q
a^{\a\b}=-\f{i}{2}\p^{[\a\g}B_\g^{\b]}-i\p^{\a\b}\Sigma~,
$$
as well as the on-shell conditions \be\Box\Sigma=0,\q
\p^{(\a\g}B_\g^{\b)}=0, \q \p^{\a\b}\kappa_{\b}^i=0~.\ee At the same
time  the fields $C^{(-2)}, \varphi^{(-3)\a}, E^{(-2)\a\b}$  are
eliminated by the choice of gauge $f^{ij}=0,$ $ \r^i_\a =0~.$

The free action for the dynamical equations (\ref{EqB}),
(\ref{EqPhi}) has been proposed in \cite{sok} \be\label{lagrPhi}
S_{TM}= \int d^6xd^8\t du \Phi^{(-3)}_\a D^{++}{\cV}^{+\a}=\int
d^6xd^8\t du \Phi D^{++}{\cV}^{(-2)}~.\ee  This action is invariant
under the above gauge transformations of the ${\cV}^{(-2)}$ together
with the gauge invariant condition for ${\Phi}$ 
($\d_\Lambda\Phi=0),$
$$\d S_{TM}=\int d^8\t du \Phi D^{++}D^{--}\Lambda= \int d^8\t du D^{++}\Phi D^{--}\Lambda=0~,$$
where the on-shell equation $D^{++}\Phi=0~$ has been used. Note also
that all the constraints (\ref{cV}), (\ref{constPhi}) for the
superfields ${\cV}^{-\a}=D^{--}{\cV}^{+\a}$ and $\Phi$ can be solved
in the anti-analytic basis of the harmonic superspace, where
$D^-_\a=-\f{\p}{\p\t^{+\a}}$:
\be{\cV}^{-\a}(x_{\bar{A}},\t^-,\t^+,u)=v^{-\a}(\t^-)+\t^{+\a}v^{(-2)}(\t^-)=(D^-)^{3\a}{\cV}^{+2}~,\ee
and
\be\Phi(x_{\bar{A}},\t^-,\t^+,u)=l(\t^-)+\t^{+\a}f^-_\a(\t^+)=(D^-)^{3\a}\Phi^{(+3)}_\a~.\ee

This BF-type action (\ref{lagrPhi}) describes two tensor multiplets
one of which acts as a Lagrange multiplier for the equations of
motion of the other multiplet: \be\label{BF} S= \int d^6x
(G^+_{\a\b}\p^{(\a\g}B_\g^{\b)}+i\psi^+_\a\p^{\a\b}k^-_\b+\phi\Box\Sigma)~.
\ee We see that the superfield $\Phi^{(-3)}_\a$ describes those
degrees of freedom, which are killed in 3-form $G_{abc}$ by the
self-duality condition. According to the work \cite{sok} the
self-dual fields do not exist off-shell on their own\footnote{It
would be interesting to quantize such a theory and study the
effective action analogously to the self-dual YM theory \cite{SK91}.
One can expect that these fields can be propagating due to quantum
corrections.}.

In the analytic subspace of the harmonic superspace the analytic
superfields \be\label{G^{++}}{\cal G}^{++}=D^+_\a\Phi
{\cV}^{+\a}+\f14\Phi D^+_\a {\cV}^{+\a},\q D^+_\a {\cal G}^{++}=0~,
\ee allow us to rewrite the action (\ref{lagrPhi}) in the form
\be\label{actT} S= \int d\zeta^{(-4)}du\{D^+_\a\Phi
D^{++}{\cV}^{+\a}+\f14\Phi D^{++}D^+_\a {\cV}^{+\a}\}~.\ee This
expression completely corresponds to the standard recipe for
constructing the superfield action in harmonic/projective superspace
(see e.g. \cite{gios}, \cite{SKD5}, \cite{project}) and will be used
below for constructing the interacting superfield action of the
vector/tensor system.

\subsection{(2,0) Tensor multiplet}
The field content of the six dimensional (2,0) tensor multiplet
consists of a self-dual 3-form curvature
$G_{(\a\b)}(x)=(\g^{abc})_{\a\b}G_{abc}$ with three on-shell degrees
of freedom, four left-handed spinors $\psi^I_\a(x)$ and five scalars
$\phi^{IJ}(x)=-\phi^{JI}(x)$ which satisfy the condition
$\O_{IJ}\phi^{IJ}=0$ \cite{FS}. All these component fields can be
encoded into the $\Omega$-traceless scalar superfield
$L^{[IJ]}(x,\theta^{I})$ (I,J=1,2,3,4; ${\bf 5}$ of $USp(4)$),
subject to the differential constraints \be\label{20L} D^K_\a
L^{IJ}-\f25 D_{\a
L}(\O^{KI}L^{LJ}-\O^{KJ}L^{LI}+\f12\O^{IJ}L^{LK})=0.\ee One can also
impose the reality condition
$\overline{L_{IJ}}=\O_{IK}\O_{JL}L^{KL}.$ The constraints on the
trace-free part of $D_\a^K L^{IJ}$ arise as a consistency condition
on the embedding an M5-brane in an eleven-dimensional superspace
\cite{HSWest}. Using the spinor derivative algebra (\ref{salg}) it
is not difficult to show that this superfield has the following $\t$
expansion \be\label{komponetL}
L^{IJ}=\phi^{IJ}+(\t^{\a[I}\psi^{J]}_\a+\f12\O^{IJ}\t^{\a K}\psi_{\a
K})+(\t^{\a [I}\t^{\b J]}+\f12\O^{IJ}\t^{\a
K}\t_K^\b)\f12G_{(\a\b)}+\ldots~.\ee The corresponding component
fields satisfy the massless equations of motion
$$\Box\phi^{IJ}=0,\q \p^{\a\b}\psi^I_{\b}=0, \q \p^{\a\g}G_{\g\b}=0.$$
The latter equation implies that the 3-form $G_{abc}$ is the curl of
a 2-form $G_{abc}=\p_{[a}B_{bc]}$, or
$G_{\a\b}=\p_{(\a\g}B^\g_{\b)}$. The gauge transformations now take
the form $B_\b^\a \rightarrow
\p^{\a\g}\Lambda_{\b\g}-\f14\d^\a_\b\p^{\g\d}\Lambda_{\g\d} .$

There are various complications in formulation of (2,0) interacting
theories with non-Abelian gauge group \cite{LP}, \cite{MMM},
\cite{Lambert}. It is still unclear whether a superfield action for
this multiplet actually exists.

\section{(2,0) tensor multiplet in D6, (1,0) harmonic superspace}
\def\theequation{4.\arabic{equation}}
\setcounter{equation}{0} In this section we are going to show that
the (2,0) tensor multiplet can be formulated in (1,0) harmonic
superspace in terms of the (1.0) tensor multiplet and
hypermultiplet.

It is easy to see that the total on-shell field contents of the
(1,0) hypermultiplet and the (1,0) tensor multiplet  exactly
coincides with one of the (2,0) tensor multiplet. Therefore it seems
natural that the dynamical theory of the (2,0) tensor multiplet can
be constructed in the (1,0) harmonic superspace in terms of the
(1,0) hypermultiplet and the (1,0) tensor multiplet. Consider a sum
of actions for hypermultiplet (\ref{actHyper}) and (1,0) tensor
multiplet (\ref{EqB}). We show that this total action possesses by
extra hidden (1,0) supersymmetry. Taking into account the manifest
(1,0) supersymmetry of the actions (\ref{actHyper}) and (\ref{EqB}),
one gets a (2,0) supersymmetry of the total action.

Let us write the above total action in the form \be\label{total}
S^{(2,0)}= S_q + S_T =\int d^6xd^8\t du \Phi
D^{++}{\cV}^{(-2)}+\f12\int d\zeta^{(-4)}du {q}^+_{A}
D^{++}q^{+A}.\ee Here $A=1,2$ is the index of the Pauli-G\"ursey
rigid $SU(2)$ symmetry. As mentioned in the previous section, there
are two ways to interpret the action $S_T$. Therefore we can define
two types of hidden supersymmetry transformations.

First, we treat the superfield $\Phi$ in action $S_T$ as Lagrangian
multiplier and ${\cV}^{(-2)}$ as the basic superfield. We define the
hidden supersymmetry transformations in the form \be\label{trans1}\d
q^+_A= (D^+)^4\epsilon^\a_A\Phi^{(-3)}_\a, \q \d {\cV}^{(-2)}=-
\epsilon^\a_A(\t^-)^{3}_\a q^{+A}, \q \d \Phi=0 ~,\ee where
$\epsilon^\a_A$ is the transformation parameter. Then, the variation
of the hypermultiplet action is \be \d S_q= \int d^6xd^8\t
du\epsilon^\a_A \Phi^{(-3)}_\a D^{++}q^{+A}~. \ee The variation of
the tensor multiplet action looks like \be \d S_T= \int d^6xd^8\t
du\Phi D^{++}\d {\cV}^{(-2)}=\int d^6x d^8\t du \Phi^{-3}_\a
D^{++}(D^+)^{3\a}\d {\cV}^{(-2)}\ee
$$ =-\int d^6x d^8\t
du \epsilon^\a_A \Phi^{(-3)}_\a D^{++}q^{+A}~.$$ We see that $\d S_q
+ \d S_T=0$.

Second, we treat the superfield ${\cV}^{(-2)}$ in action $S_T$ as
the Lagrangian multiplier and the superfield $\Phi$ as the basic
superfield. In this case we define the hidden supersymmetry
transformations in the form \be\label{trans2}\d q^+_A=
(D^+)^4\epsilon^\a_AD^{-}_\a {\cV}^{(-2)}, \q \d {\cV}^{(-2)}=0, \q
\d\Phi=-\epsilon^{\a A}D^-_\a q^{+}_{A} ~\ee Then \be \d S_q= \int
d^6xd^8\t du\epsilon^\a_A D^-_\a {\cV}^{(-2)}D^{++}q^{+A},~\ee and
\be \d S_T= -\int d^6x d^8\t du\epsilon^{\a A}D^{++}q^{+}_{A}D^-_\a
{\cV}^{(-2)}= \int d^6xd^8\t du\epsilon^\a_Aq^{+A} D^{++}D^-_\a
{\cV}^{(-2)}~. \ee We see again that $\d S_q+\d S_T=0.$

As a result we have constructed the free action for the (2,0) tensor
multiplet in the (1,0) harmonic superspace in terms of the (1,0)
hypermultiplet and the (1,0) tensor multiplet. This action is
invariant under the manifest (1,0) supersymmetry transformations and
under the hidden (1,0) supersymmetry transformations.

It is interesting to study whether the supersymmetry algebra is
closed. Let us begin with formulation on the base of superfields
$q^{+}$ and ${\cV}^{(-2)}$. The transformation laws for these
superfields are given by (\ref{trans1}).
 Then it is not
difficult to obtain that \be[\d_2,\d_1]\Phi=2i\epsilon^{\b
A}_1\epsilon^\a_{2 A}\p_{\a\b}\Phi.\ee Here we have used the
identity $(D^+_\a D^-_\b+D^-_\a D^+_\b)\Phi=0$ which follows from
the constraints (\ref{constPhi}). For the hypermultiplet we have \be
[\d_2,\d_1]q^+_A=2i\epsilon^{\a B}_{1
}\epsilon^\b_{2B}\p_{\a\b}q^{+}_A~.\ee We see that the algebra of
the hidden supersymmetry transformations is closed. An analogous
consideration can be carried out for the formulation with basic
superfields $q^{+}$ and $\Phi$. The corresponding algebra is also
closed.

\section{The interacting D6 (1,0) vector and tensor multiplets\\ in harmonic superspace}

\def\theequation{5.\arabic{equation}}
\setcounter{equation}{0}

\subsection{Non-Abelian vector/tensor system}
In this subsection we will briefly mention the general non-Abelian
couplings of vectors and antisymmetric $p$-form fields in six
dimensions following \cite{SSWW}. The (1,0) superconformal $6D$
field theory of \cite{SSWW} (vector/tensor system) describes a
hierarchy of non-Abelian scalar, vector and tensor fields $\{\phi^I,
A_a^r, Y^{ij\  r}, B_{ab}^I, C_{abc\ r}, C_{abcd\ A}\}$ and their
supersymmetric partners that label by the indices $r=1,\l, n_V$ and
$I=1,\l,n_T.$ To label the $C_{abc\ r}$ field a dual index $r$ is
used since the vector fields are dynamically dual to the
antisymmetric three-form tensors. The full non-Abelian field
strengths of vector and two-form gauge potentials are given as \be
{\cF}^r_{ab}=\p_{[a}A^r_{b]}-f_{st}^{\ \ r}A^s_a A^t_b +h^r_I
B^I_{ab}, \ee\label{calF}
$${\cH}^I_{abc}= \f12\cD_{[a}B_{bc]}+d^I_{rs}A^r_{[a}\p_b A^s_{c]}-
\f13f_{pq}^{\ \ s}d^I_{rs}A^r_{[a}A_b^pA_{c]}^q+g^{Ir}C_{abc\ r}.$$
Here  $f_{[st]}^{\ \ r}$ are the structure constants, $d^I_{(rs)}$
are the $d$-symbols, defining the Chern-Simons couplings, and
$h^r_I, g^{Ir}$ are the covariantly constant tensors, defining the
general St\"uckelberg-type couplings among forms of different
degrees. The existence of the non-degenerate Lorentz-type metric
$\eta_{IJ}$, so that $h^r_I=\eta_{IJ}g^{Jr}\equiv g^r_I$,
$b_{Irs}=2\eta_{IJ}d^J_{rs}\equiv d_{Irs}$ is also assumed. The
covariant derivatives are defined as $\cD_m=\p_m-A^r_m X_r$ with the
gauge generators $X_r$ acting on the different fields as follows:
$X_r\cdot \Lambda^s\equiv -(X_r)_t^s\Lambda^t~,$ $X_r\cdot
\Lambda^I\equiv -(X_r)_J^I\Lambda^J~.$ The covariance of the field
strengths (\ref{calF}) requires that the gauge group generators in
the various representations should have the form
$$(X_r)_s^t=-f_{rs}^{\ \ t}+g^t_I d^I_{rs}, \q (X_r)_I^J=2d^J_{rs}g^s_I-g^{Js}d_{Isr},$$
in terms of the invariant tensors parameterizing the system (see the
details in \cite{SSWW}). The field strengths (\ref{calF}) are
defined such that they transform covariantly under the set of
non-Abelian gauge transformations \be\label{trA}\d
A_m=\cD_m\Lambda^r -h_I^r\lambda^I_m,\ee
$$\d B_{mn}^I=\cD_{[m}\Lambda^I_{n]}-2d^I_{rs}(\Lambda^r{\cF}^s_{mn}-
\f12A^r_{[m}\d A^s_{n]})-g^{Ir}\Lambda_{mn\ r}.$$

The superspace realization of the tensor hierarchy has been
developed in the paper \cite{IB} in conventional 6D, (1,0)
superspace. In the next subsection we are going to consider the
generalized Bianchi identities from \cite{IB} for the superfield
vector/tensor system, reformulate them in the harmonic superspace
and study the consistency conditions for the generalized Bianchi
identities. For further use, it is convenient to introduce the
generalized superfield strength \be\label{genstrength} {\cW}^{i\a\
r}=W^{i\a \ r}+g_I^r{\cal V}^{i\a \ I}, \ee where the $W^{i\a \ r}$
is the superfield strength of the super Yang-Mill theory (defined in
subsection 3.2) and ${\cal V}^{i\a \ I}$ is the superfield of the
tensor multiplet (defined in subsection 3.3), and write the
generalized Bianchi identities in its terms. Then one can see that
the conventional strength $F_{mn}$ of the vector multiplet and the
conventional strength $B_{mn}$ of the tensor multiplet enter into
${\cW}^{i\a\ r}$ in the form $F_{mn} + gB_{mn}.$

Using the generalized Bianchi identities and their consistency
conditions we will formulate the superfield action for vector/tensor
system and find its component form.

\subsection{A harmonic superspace description of non-Abelian vector/tensor system}
In this subsection we are going to formulate the superfield version
of non-Abelian vector/tensor system using the harmonic superspace
technique. A complete set of the constraints on superfield strengths
of the $p$-form potentials has been proposed in \cite{IB} in
conventional 6D, (1,0) superspace. Our aim is to reformulate these
constraints in harmonic superspace and study their consistency
conditions.

First of all, the SYM constraint ${\cF}_{\a\b}^{ij\ \ r}=0$ is not
deformed, therefore we can use a harmonic superfield technique. Then
we consider the dimension 2 component of the generalized Bianchi
identities
 \be
 (\g^a)_{(\a\d}\cD^{(j}_{\b)}{\cW}^{i)\d \ r}-2\ve^{ij}(\g^b)_{\a\b}{\cF}_{ab}^r
=\f12\ve^{ij}\g^a_{\a\b}\Phi^Ig_I^r.\ee This relation leads to the
covariant derivatives of generalized superfield strength
(\ref{genstrength}) in the form \be   \cD^i_\a {\cW}^{j\b\
r}=\d_\a^\b( {\cY}^{ij\ r}+\f12\ve^{ij}\Phi^Ig_I^r)
+\f12\ve^{ij}(\g_{ab})_\a^\b {\cF}^r_{ab}~.\ee This equation is
equivalent to the following set of relations \be\label{relat}
\cD_\a^{(i}{\cW}^{j)\b}=\f14\d^\b_\a \cD_\g^{(i}{\cW}^{j)\g}, \q
{\cY}^{ij}=\f18\cD_\a^{(i}{\cW}^{j)\a}, \q
\Phi^Ig_I^r=\f14\cD_{i\a}{\cW}^{i\a \ r},\q
{\cF}_{ab}=-\f{1}{8}(\g_{ab})_\a^\b \cD_{i\b}{\cW}^{i\a}~.\ee

We turn now to harmonic superspace formulation. In terms of the
harmonic superfields, the relations  (\ref{relat}) take the form \be
\cD^+_\a {\cW}^{+\b\ r}=\d_\a^\b {\cY}^{++\ r},\q\q \cD^-_\a
{\cW}^{+\b\ r}=\d_\a^\b ({\cY}^{+-\ r}+\f12\Phi^Ig_I^r)+\f12
{\cF}_\a^{\b\ r}\ee
$$\cD^-_\a {\cW}^{+\b\ r}-\cD^+_\a {\cW}^{-\b\ r}=\d_\a^\b\Phi^Ig_I^r +
{\cF}_\a^{\b\ r},\q \f14(\cD^-_\a {\cW}^{+\a\ r}-\cD^+_\a {\cW}^{-\a\ r})=\Phi^Ig_I^r~.$$

Consider the dimension 5/2 component of the generalized Bianchi
identities
\be\label{Psi}\cD^i_\a{\cF}^r_{ab}+i(\g_{[a})_{\a\d}\cD_{b]}{\cW}^{i\d\
r} =i(\g_{ab})_\a^\b\Psi^{i I}_\b g_I^r~.\ee
 It
yields $ 3\Psi^{iI}_\a g_I^r=\f{i}{10}\cD^i_\b {\cF}_\a^{\ \b\
r}+\cD_{\a\b}{\cW}^{i\b\ r}~.$ In addition, the above deformed identity
determines the transformation law for the potential of 2-forms
$$\d B_{ab}^I=i\epsilon_i\g_{ab}\Psi^{iI}~.$$
The selfconsistency conditions
$\{\cD^i_\a,\cD^j_\b\}{\cW}^{k\d}=-2i\ve^{ij}\cD_{\a\b}{\cW}^{k\d}$
leads to \be \cD_\a^i\Phi^I=2i\Psi^{iI}_\a, \q i\cD_{\a\b}{\cW}^{\b
i\ r}=-\f{1}{3}\cD_{\a l}{\cY}^{li\ r}+\cD^i_\a\Phi^Ig_I^r~,\ee
\be
\cD_\a^k {\cY}^{ij\ r}=-i\ve^{k(i}(\cD_{\a\b}{\cW}^{\b j)\
r}-2\Psi_\a^{j)I}g_I^r)~.\ee By rewriting the above relations in
terms of harmonic projection, one gets  $\cD^+_\a {\cY}^{++\ r}=0, $
\be\label{DY} \cD_\a^- {\cY}^{++\ r}=-2i(\cD_{\a\b}{\cW}^{+\b\ r}-
2\Psi^{+I}_\a g^r_I), \q \cD_\a^\pm {\cY}^{+-\ r}=\pm
i(\cD_{\a\b}{\cW}^{\pm\b\ r}- 2\Psi^{\pm I}_\a g_I^r )~.\ee Acting
on the relation $\cD_\a^i\Phi^I=2i\Psi^{iI}_\a$ by the spinor
derivative, one obtains \be \cD_a\Phi^I=\f{1}{4} \cD_{\a
i}{\g}_a^{\a\b}\Psi_\b^{iI}~. \ee

The dimension 3 component of the generalized Bianchi identities is
\be \cD_{[a}{\cF}_{bc]}={\cH}^I_{abc}g_I^r, \q
{\cH}_{abc}=({\cH}^{(+)}+{\cH}^{(-)})_{abc},\ee In the spinor
notations it has the form \be\label{H}
\f12\cD_{(\a\d}\tilde{{\cF}}^{\d \ r}_{\
\b)}=\f13{\cH}^{(-)I}_{\a\b}g_I^r,\q \f12\cD^{(\a\d}{{\cF}}^{\b)\
r}_{\d}=\f13{\cH}^{{(+)}\a\b I}g_I^r~.\ee  The symmetric in $(i,j)$
parts of the equations (\ref{Psi}) have the form
$$ \cD^{(i}_{(\a}(\g_{ab})_{\b)}^\d\Psi^{j) I}_\d=2id^I_{rs}{\cW}^{\r i\ r}\g^{[a}_{\r(\a}\g^{b]}_{\b)\d}{\cW}^{\d j\ s}, \q \f12\cD^{(i}_\a\g_a^{\a\b}\Psi^{j)I}_\b=
2i{\cW}^{i\ r}\g_a {\cW}^{j\ s}d_{rs}^I~.$$ The antisymmetric in
$(i,j)$ parts of the same equations have the form \be {\cH}^{(+)
I}_{abc}=\f{i}{4}d^I_{rs}{\cW}^{\a\ r}_i\g^{abc}_{\a\b}{\cW}^{i\b \
s},\q {\cH}^{(-)I}_{abc}=
\f{1}{8}\cD_{i\a}\g^{\a\b}_{abc}\Psi^{iI}_\b~.\footnote{An important
feature of these equations is that the anti-self-dual part of the
field strength ${\cH}$ is fixed in terms of the dynamical vector
multiplet.}\ee The above equations for symmetric and antisymmetric
parts together imply \be \cD_\a^i\Psi^{j
I}_\b=-\f{1}{2}\ve^{ij}\cD_{\a\b}\Phi^I-\f{1}{12}\ve^{ij}\g^{abc}_{\a\b}{\cH}^{(-)
I}_{abc} +i\ve_{\a\b\g\d}{\cW}^{i\g \ r} {\cW}^{j\d\ s}d^I_{rs}~.\ee
In terms of the harmonic superfields these relations take the form
\be\label{DP}\cD_\a^\mp\Psi^{\pm
I}_\b=\mp\f{1}{2}\cD_{\a\b}\Phi^{I}\mp\f{1}{12}\g^{abc}_{\a\b}{\cH}^{(-)I}_{abc}
+i\ve_{\a\b\g\d}{\cW}^{-\g r} {\cW}^{+\d s}d^I_{rs}~,\ee
$$\cD^\pm_\a \cD^\pm_\b\Phi^I=-2\ve_{\a\b\g\d}{\cW}^{\pm\g\ s}{\cW}^{\pm\d\ r}d^I_{sr}~,
$$
$$ \cD^-_{\a} \cD^-_{\b}{\cY}^{++\ r}=-i(\cD_{\a\b}\Phi^Ig_I^r+\f13\g^{(3)}_{\a\b}{\cH}^{(-)\ I}_{(3)}g_I^r-\cD_{\b\g}{\cF}^{\g\ r}_{\ \a}-2\cD_{\a\b}{\cY}^{+-\ r})~.$$

The spinor derivative of the 3-rank tensor superfield ${\cH}^{
I}_{abc}$ is \be \cD_\a^i {\cH}^I_{abc}=i\g^{[a}_{\a\b}{\cW}^{i\b\
r}{\cF}^{bc]\ s}d^I_{sr}
+\f{i}{2}\cD^{[a}(\g^{bc]})_\a^\b\Psi^{Ii}_\b
-i\g^{abc}_{\a\b}{\cW}^{i\b\ s}\Phi^J d_{Jsr}g^{Ir}~.\ee This
relation also determines  the transformation law of the 3-form
potential
$$\d C_{abc \ r}=-i\epsilon_i\g_{abc}{\cW}^{i\ s}\Phi^J d_{Jsr}.$$
The corresponding degrees of freedom are not dynamic since the
generalized 4-form field strength satisfies the duality conditions
\be-\f{1}{4!}\ve^{abcdef}{\cH}_{abcd\ r}=({\cF}^{ef\
s}\Phi^I+i{\cW}^{i\ s}\g^{ef}\Psi^I_i)d_{I\ rs}.\ee

As shown in the paper \cite{IB}, all other relations among the main
superfield strengths and the equations of motion can be derived from
the following relations
\be({\cY}^{ij\ s}\Phi^I-2i{\cW}^{(i \ s}\Psi^{j)
I})d_{I\ rs}=0~,\ee
$$d_{I rs}\{\Phi^I\cD_{\a\b} {\cW}^{\b s}_i +\f{1}{12}{\cH}^{(-)I}_{\a\b}{\cW}^{\b \ s}_i+\f12\cD_{\a\b}\Phi^I {\cW}^{\b\ s}_i+\f12{\cF}_\a^{\ \b s}\Psi^I_{\b i}
+{\cY}_{ij}^s\Psi^{I\ j}_{\a}\}=$$
$$+\f12\Psi^I_{\a i}\Phi^J(4g_I^s d_{Jrs}-g_J^s d_{Irs})+\f{2i}{3}\ve_{\a\b\g\d}{\cW}^{\b js}{\cW}^{\g u}_j{\cW}^{\d v}_i d^I_{rs}d_{I uv},$$
$$\cD_e(\Phi^I{\cF}^{ea\ s}+i{\cW}^{i\ s}\g^{ea}\Psi^I_i)d_{Irs}+\f16\ve^{abcdef}{\cF}^s_{bc}{\cH}^I_{def}d_{Irs}$$
$$+(-\f14\Phi^{[I}\cD^a\Phi^{J]}+\f{i}{2}\Psi^{iI}\tilde\g^a\Psi^J_i)X_{rIJ}-\f{i}{2}\Phi^I {\cW}^{is}\g^a {\cW}^t_i (X_r)^u_{\ [s}d_{I\ t]u}=0.$$
They lead to the Dirac equation for the fermions of the tensor
multiplet \be \cD^{\a\b}\Psi^{iI}_\b=-{\cY}^{ij\ r}{\cW}^{\a\
s}_jd_{rs}^I+\f12 {\cW}^{i\a\ r}\Phi^J(4d_{J
rs}g^{Is}-g^s_Jd^I_{sr})-\f12 {\cW}^{i\b\ r}{\cF}_\b^{\ \a\
s}d^I_{rs}~,\ee to scalar superfield equation of motion  \be
\Box\Phi^I=d^I_{rs}({\cF}^r_{ab}{\cF}^{ab\ s}-{\cY}^r_{ij}{\cY}^{ij\
s}-i{\cW}^{\a \ r}_i\cD_{\a\b}{\cW}^{i\b\
s})+\f32\Phi^Jg_J^r\Phi^Kg^s_Kd^I_{rs}\ee
$$ -i{\cW}^{i\a \
r}\Psi_{i\a}^{\ J}(4g_J^sd^I_{rs}-g^{Is}d_{J\ rs})~,$$ and to the
second order equations for the 3-form field strength ${\cH}^I_{abc}$
$$\cD^c{\cH}^I_{abc}=-\f14\ve_{abcdef}{\cF}^{cd\ r}{\cF}^{ef\ s}d^I_{rs}+{\cF}^r_{ab}\Phi^Jd_{Jrs}g^{rI}+i{\cW}^r_i\g_{abc}\cD^c {\cW}^{i\ s}d^I_{rs}
-i{\cW}^{i\ r}\g_{ab}\Psi^J_i g_J^s d^I_{sr}.$$

These equations of motion allow us to construct a component action
of the theory in the form
$$S=\int d^6x\{\f12\cD^{a}\Phi_I\cD_{a}\Phi^I+\Phi_I d^I_{rs}( -Y^{ij\ r}Y_{ij\ s}
+{\cF}^{ab\ r}{\cF}^s_{ab}-iW_i^{\a\ r} \cD_{\a\b}W^{i\b\ s}
+\f12\Phi^J \Phi^Lg_J^r g_L^s)
$$
$$+\Phi_I iW^{\a\ r}_i\Psi^{i\ J}_\a(4g_J^sd^I_{rs}-g^{Is}d_{J\ rs})+i\Psi^i_{\a\ I}\cD^{\a\b}\Psi^I_{i\b}-2i\Psi_{i\a\ I}W^{\a\ r}_jY^{ij\ s}d^I_{rs}
+i\Psi_{\a\ I}^i W^{\b\ r}_i{\cF}_\b^{\ \a\ s}d^I_{rs}
$$
\be\label{component}+\f{1}{6}{\cH}_I^{abc}{\cH}^I_{abc}+
\f{i}{12}W^{i\a\ r }\g^{abc}_{\a\b}W_i^{\b\
s}{\cH}^I_{abc}d_{Irs}-\f{1}{3}\ve_{\a\b\g\d}W^{\a\ r}_j W^{i\b \
s}W^{\g\ u}_i W^{j\d\ v} d^I_{rs}d_{I uv}\}~.\ee Here we assume the
existence of the non-degenerate symmetric metric
$\eta_{IJ}=g_{Ir}g^r_J~.$ Also the first components of the
superfields are denoted the same way as the corresponding
superfields, e.g. ${\cW}^{\a i}|_{\t=0}=W^{\a i}$,
${\cY}^{ij}|_{\t=0}=Y^{ij}, \l .$ The action (\ref{Scomp}) coincides
(up to field redefinition) with the component action of the
superconformal vector/tensor system constructed in the papers
\cite{SSWW},\cite{BSPr}.

The other set of equations  given in \cite{IB} includes
supersymmetrizations of the Hodge-duality relations between the
3-form potential and the non-Abelian vectors and scalar-4-forms
relations. All the relations of this subsection are used in the next
subsection for finding the component form of superfield action of
the vector/tensor system.

\subsection{Superfield Lagrangian formulation of the vector/tensor system}

In this subsection we are going to propose the superfield action for
the non-Abelian vector/tensor system in harmonic superspace and find
its component form.

Let us introduce the superfield \be\label{Y} \Upsilon^I
=\Phi^I+\f{1}{2}d^I_{rs}(D^+_\a V^{-- r}{\cal W}^{+{\alpha}s}+2V^{--
r}{\cal Y}^{++ s})~,\ee where \be\label{y} {\cal Y}^{++ s} = Y^{++
s} +\f{1}{4}D^{+}_{\alpha}{\cal V}^{+{\alpha} s}. \ee One should
remember that the $Y^{++}$ is defined in subsection 3.2 and the
${\cal V}^{+{\alpha} s}$ is defined in subsection 3.3. The
expression $\Upsilon^I$ (\ref{Y}) is the only extension of $\Phi$
preserving linearity, $ D^+_\a D^+_\b\Upsilon=0,$ i.e the
$\Upsilon^I$ is a linear superfield. By using the superfield
(\ref{Y}), one can define the superfield action in harmonic
superspace as follows \be\label{S} S = \int d\zeta^{(-4)} dug_{I
r}\{\Upsilon^I \cD^{++}{\cal Y}^{++\ r}+D^+_\a\Upsilon^I
\cD^{++}{\cal W}^{+{\alpha}r}\}~.\ee The invariant tensor $g_{I r }$
has already been defined in \cite{SSWW}. The integrand of the
expression (\ref{S}) is the only (up to common coefficient) analytic
superfield constructed from $\Upsilon^I$, ${\cal Y}^{++}$, ${\cal
W}^{+{\alpha}}$ and contains no higher derivatives of ${\cal
D}^{++}$, $D^{+}_{\alpha}$. The action (\ref{S}) depends both on
superfields of the vector multiplet and superfield $\Phi$
responsible for the tensor multiplet. If $\Phi$ is a constant
$1/f^2$, this action takes the form (\ref{actZ}) of SYM action
$S\sim \f{1}{f^2}\int d^6x d^8\t du V^{++}V^{--}$. Besides, the
proposed action possesses supersymmetry and  gauge symmetries of
vector/tensor system.

The action (\ref{S}) is the natural generalization of the free
action (\ref{actT}). Indeed if we put in (\ref{Y}) $\Upsilon = \Phi$
and use the relations (\ref{y}), (\ref{genstrength}) and the
identities $D^{++}Y^{++}=0$, $D^{++}W^{+{\alpha}}=0$, one gets the
action (\ref{actT}). Thus, the action (\ref{S}) is the only possible
superfield action for the non-Abelian vector/tensor system which has
the free action (\ref{actT}) in the Abelian limit.

Now we will derive the component form of the action (\ref{S}). For
simplicity we are only going to consider the Abelian case. By
integrating over the anticommuting coordinates, one gets
\be\label{Scomp} S = \f18 \int d^{6}x_A du( \cD^-)^4g_{I\
r}\{\Upsilon^I \cD^{++}{\cal Y}^{++\ r}+D^+_\a\Upsilon^I
\cD^{++}{\cal W}^{+\a \ r}\}|_{\t=0}~.\ee Further we act by the
derivatives $D^-_\a$ and put all the theta's equal to zero. Then act
by the harmonic derivative $\p^{++}$. After the cumbersome enough
calculations\footnote{The intermediate calculations are given in
Appendix B with use of the relations from the subsection 5.2.}, we
obtain all the functional structures which are present in the
component action of the vector/tensor system
(\ref{component})\footnote{Derivation of a component action in a
non-Abelian case requires an additional study.}.


\section{One-loop effective action in the hypermultiplet\\ theory}
\def\theequation{6.\arabic{equation}}
\setcounter{equation}{0}

In this section we will consider a calculation of the superfield
quantum effective action in the hypermultiplet theory coupled to the
external field of vector/tensor system. We will show that the (1,0)
super Yang-Mills action (\ref{actZ}), the vector/tensor multiplet
action (\ref{actT}) or (\ref{Scomp}) and higher derivative vector
multiplet action \cite{ISZ} are generated as the divergent parts of
the effective action. For simplicity we will assume that the
background is Abelian.

The classical conformal invariant action for a massless
hypermultiplet of canonical dimension 2 coupled to a background 6D
$\cN=(1,0)$ vector/tensor system is written as \be\label{hiper}
S=-\f12\int du d\zeta^{(-4)}q^{+A}\cD^{++}q^+_A=-\int du
d\zeta^{(-4)}\tilde{q}^{+}\cD^{++}q^+~,\ee with
$\cD^{++}=D^{++}+gV^{++}$ the analyticity-preserving covariant
derivative and $V^{++}$ the analytic potential. We want to emphasize
that the superfield $V^{++}$ here is not one for pure vector
multiplet, the superfield strengths, involving the superfield
$V^{++}$, obey the Bianchi identities which contain the superfield
$\Phi$ related to tensor multiplet (see subsection 5.2). As a result
the action (\ref{hiper}) describes interaction of hypermultiplet
with vector/tensor system. The dynamical variable $q^+$ is a
covariantly analytic superfield and $\tilde{q}^+$ is the conjugate
of $q^+$ with respect to the analyticity preserving conjugation
\cite{gios} $ q^{A+}=\epsilon^{AB}q^+_B=\widetilde{(q^+_A)},\q
q^+_A=(q^+,-\tilde{q}^+)~.$

The hypermultiplet effective action $\Gamma$ is defined by
\be\label{EAhyper} e^{i\Gamma[V^{++}]}=\int\cD q^+\cD \tilde{q}^+
\exp(-i\int d\zeta^{(-4)}\tilde{q}^+\cD^{++}q^+)~.\ee The expression
(\ref{EAhyper}) yields
\be\label{Tr}\Gamma[V^{++}]=i\Tr\ln\cD^{++}=-i\Tr\ln G^{(1,1)}.\ee
Here
$G^{(1,1)}(\zeta_1,u_1|\zeta_2,u_2)=<\tilde{q}^{+}(\zeta_1,u_1){q}^{+}(\zeta_2,u_2)>$
is the superfield Green function in the $\tau$-frame. This Green
function is analytic with respect to both arguments and satisfies
the equation \be\label{eqG}
\cD_1^{++}G_\tau^{(1,1)}(1|2)=\d_A^{(3,1)}(1|2)~.\ee Here
$\d_A^{(3,1)}(1|2)$ is the appropriate covariantly analytic
delta-function
\be\d_A^{(q,4-q)}=(D^+_2)^4\d^{14}(z_1-z_2)\d^{(q,-q)}(u_1,u_2)=(D^+_1)^4\d^{14}(z_1-z_2)\d^{(q-4,4-q)}(u_1,u_2).\ee
The formal solution to this equation can be found analogously to
four-dimensional case \cite{graph}, \cite{HEA}, \cite{KUZ} and looks
\footnote{As well as in the work \cite{HEA} we will act by the
operator $(\cD_1^{--})^2$  on both sides of (\ref{eqG})
$$\cD^{++}_1(\cD_1^{--})^2
G^{(1,1)}(1|2)=(\cD_1^{--})^2\d_A^{(3,1)}(1|2)=\cD^{++}_1
2(\cD^+_2)^4\f{\d^{14}(z_1-z_2)}{(u^+_1u^+_2)^3}~.$$ Now, since the
equation $D^{++}f^{-|q|} = 0$ has only the trivial solution
$f^{-|q|} = 0$, after the action of the operator $(\cD^+_1)^4$ we
obtain:
$$(\cD^+_1)^4(\cD^{--}_1)^2G^{(1,1)}(1|2)=-8\sB G^{(1,1)}(1|2)=2(\cD^+_1)^4(\cD^+_2)^4\f{\d^{14}(z_1-z_2)}
{(u^+_1u^+_2)^3}~.$$} like \be\label{GREEN}
G_\tau^{(1,1)}(1|2)=-\f{1}{4\sB}(\cD^+_1)^4(\cD^+_2)^4\d^{14}(z_1-z_2)\f{1}{(u^+_1u^+_2)^3},\ee
where $1/(u^+_1u^+_2)^3$  is a special harmonic distribution. In Eq.
(\ref{GREEN}) the $\stackrel{\frown}{\Box}$ is the covariantly
analytic d’Alembertian ($[D^+_\a,\sB]=0$) which arises when
$(\cD^+)^4(\cD^{--})^2$  acts on the analytical superfield and has
the form \be\label{sBox}
\sB=-\f18(\cD^+)^4(\cD^{--})^2|=\cD_a\cD^a+{\cW}^{+\a}\cD^-_\a
+Y^{++}\cD^{--}-\f{1}{4}(\cD^-_\a {\cW}^{+\a})-\f12\Phi
~. \ee Pay attention to the fact that the term $\Phi$ in the above
relation is responsible  for the tensor multiplet contribution. Like
in four- and five-dimensional cases \cite{KUZ} one can obtain the
useful identity
\be\label{Delta}(\cD^+_1)^4(\cD^+_2)^4\f{1}{(u^+_1u^+_2)^3}=(\cD^+_1)^4
\{(u^+_1u^+_2)(\cD^-_1)^4 -(u^-_1u^+_2)\Delta^{--}
-4\sB\f{(u^-_1u^+_2)^2}{(u^+_1u^+_2)}\}~.\ee Here
$$\Delta^{--}=i\cD^{\a\b}\cD^-_\a\cD^-_\b+4{\cW}^{-\a}\cD^-_\a-(D^-_\a {\cW}^{-\a})~.$$
This identity is used later for computing the effective action.

The  definition (\ref{Tr}) of the one-loop effective action is
purely formal. The actual evaluation of the effective action can be
done in various ways (see e.g. \cite{HEA}, \cite{KUZ}). Further we
will follow \cite{KUZ} and use the relation
\be\label{Gamma}\Gamma(V)=\Gamma_{y=0}+\int_0^1 dy\p_y\Gamma(y
V)=-i\Tr\int_0^1 dy(V^{++}G^{(1,1)}(y)),\ee where \be
\Tr(V^{++}G^{(1,1)})=\int du_1
d\zeta_{1}^{(-4)}V^{++}(1)G^{(1,1)}(1|2)|_{1=2}.\ee Here
$G^{(1,1)}(yV)$ means the Green function depending on the superfield
$yV^{++}$.

The effective action in local approximation is represented as a
series in powers of the background fields and their derivatives.
Further we will consider the calculation of the effective action on
the base of the superfield proper-time technique.

It is obvious that the leading non-vanishing contribution on the
diagonal $(z_1, u_1)=(z_2, u_2)$ of the two-point function \be
-\f14\int du_1
d\zeta^{(-4)}V^{++}_1\f{1}{\stackrel{\frown}{\Box}_1}(\cD_1^+)^4(\cD^+_2)^4\delta^{14}(z_1-z_2)
\f{1}{(u^+_1u^+_2)^3}|_{1=2},\ee arises when $\cD_1^{--}$ from
$\sB_1$ hits on $(u^+_1u^+_2)|_{u_1=u_2}$ and in addition at least
eight spinor derivatives acting on the Grassmann delta-function are
required to produce a non-vanishing result,
$(\cD^-)^4(\cD^+)^4\d^8(\t_1-\t_2)|_{\t_1=\t_2}=1$. On the right
hand side of (\ref{Delta}) the third  term contains a harmonic
distribution which is singular at coincident points. However, this
singular terms does not contribute to $\Gamma^{(1)}$ in the leading
approximation since there is no necessary degree of $D^-_\a~.$

In the framework of the proper-time technique, the inverse operator
$\f{1}{\stackrel{\frown}{\Box}}$ is defined as follows
\be\label{inverse}-\f{1}{\stackrel{\frown}{\Box}}=\int_0^\infty
d(is) e^{is\stackrel{\frown}{\Box}}. \ee To avoid the divergences on
the intermediate steps it is necessary to introduce a
regularization. We will use a variant of dimensional regularization
(so called $\omega$ -regularization) accommodative for
regularization of the proper-time integral (see e.g. \cite{BK}). The
$\omega$-regularized version of the relation (\ref{inverse}) is
\be\label{omegainverse}-(\f{1}{\stackrel{\frown}{\Box}})_{reg}=\int_0^\infty
d(is)(is\mu^2)^{\omega} e^{is\stackrel{\frown}{\Box}}, \ee where
$\omega$ tends to zero after renormalization and $\mu$ is an
arbitrary parameter of mass dimension. Taking into account the
relation (\ref{omegainverse}) and the relations (\ref{GREEN}),
(\ref{Gamma}) one gets for effective action \be\label{omega}
\f{1}{4}\int du_{1}d\zeta_{1}^{(-4)}V^{++}(1)\int_0^\infty
d(is)(is\mu^2)^{\omega}
e^{is\stackrel{\frown}{\Box}_1}(\cD_1^+)^4(\cD^+_2)^4\f{1}{(u^+_1u^+_2)^3}\delta^{14}(z_1-z_2)|_{1=2}.\ee
Here
$\delta^{14}(z_1-z_2)=\delta^6(x_1-x_2)\delta^4(\theta^{+}_1-\theta^{+}_2)\delta^4(\theta^{-}_1-\theta^{-}_2).$
We use now the representation of the delta function
$$\d^{14}(z_1-z_2)=\int\f{d^6p}{(2\pi)^6}e^{ip_a\r^a}\d^8(\r^{\a}_i),$$
where
$$\r^a=(x_1-x_2)^a-2i(\t^+_1-\t^+_2)\g^a\t^-,\q \r^{\a i}=(\t_1-\t_2)^{\a i},$$
and $i=+,-$. In the expression (\ref{omega}) we commute the exponent
$\exp{ip_a\r^a}$ through all the operator factors to the left and
then use the coincidence limit. This yields to
$e^{is\stackrel{\frown}{\Box}_1(X)}\cdot \d^8(\t_1-\t_2)$ where
$X_a=\cD_a+ip_a, \q X^-_\a=\cD^-_\a+2p_{\a\b}\r^{-\b}.$ In order to
get the expansion of effective action in background fields and their
derivatives we should expand $e^{is\stackrel{\frown}{\Box}_1(X)}$
and calculate the momentum integrals. All these integrals have the
standard Gauss form.

We will now concentrate on calculating the divergent part of the
effective action. In the regularization scheme under consideration,
the divergences mean the pole terms of the form $\f{1}{\omega}.$ By
expanding the $e^{is\stackrel{\frown}{\Box}_1(X)}$ in the
(\ref{omega}) and leaving only the terms relating to divergences one
gets \be\label{integral}
e^{is\stackrel{\frown}{\Box}_1}(u_1^+u_2^+)(\cD_1^+)^4(\cD^-_1)^4\delta^{8}(\t_1-\t_2)|_{1=2}=
-\int_0^\infty\f{d(is)}{(is)^3}(is\mu^2)^\omega e^{-ism^2}\{is
Y^{++}+\f{(is)^2}{2}[\sB, Y^{++}]\}.\ee Here $m^2 = \Phi$. By
calculating the proper-time integral and extracting the pole terms
one gets for the right hand side of the above expression
\be\f{1}{\omega}m^2Y^{++}-\f{1}{2\omega}\Box
Y^{++}-\f{1}{2\omega}{\cW}^{+\a}D^-_\a Y^{++}~.\ee By using  the
conditions (\ref{DY}), we obtain finally the divergent part of the
effective action in the form \be\label{div}
\f{1}{4(4\pi)^3\omega}\int dud\zeta^{(-4)}V^{++}(\cD^+_\a\Phi
{\cW}^{+\a}+\Phi Y^{++})- \f{1}{2(8\pi)^3\omega}S_{ISZ}~.\ee In
principle there is also the term $ \f{1}{\omega}\int
dud\zeta^{(-4)}V^{++}{\cW}^{+\a}i\cD_{\a\b}{\cW}^{+\b}$. However it
is cancelled out with the corresponding term
$-i\cD^{\a\b}\cD^-_\a\cD^-_\b\f{(is)^2}{2}
 {\cW}^{+\g}\cD^-_\g{\cW}^{+\d}\cD^-_\d$
from the second order expansion $\Delta^{--}e^{is\sB}$.

The divergent part of the effective action contains two
contributions. The first of them is the part of the Abelian action
(\ref{S}) of the vector/tensor system proposed in section 5. If we
consider a sum of the action (\ref{S}) and first term in
(\ref{div}), we will see that this term from (\ref{div}) determines
a renormalization of the coupling constant in the
(\ref{S})\footnote{Coupling constant g is defined through the
covariant derivative ${\cal D}^{++}=D^{++} + g V^{++}$.} The second
contribution is the Ivanov-Smilga-Zupnik Abelian higher derivative
action of the vector multiplet \cite{ISZ}\footnote{The action
(\ref{ISZ}) can be written in the different superfield forms
$$S_{ISZ}=\int
 dud\zeta^{(-4)}V^{++}\Box Y^{++}=-\f18\int d^6xd^8\t V^{++}(D^{--})^2
Y^{++}=-\f18\int d^6xd^8\t V^{--}[D^{++},D^{--}] Y^{++}.$$}
\be\label{ISZ} S_{ISZ}=\f{1}{2}\int dud\zeta^{(-4)} Y^{++}
Y^{++}.\ee One should emphasize once more that we have considered
only the divergent parts of the effective action. Of course, the
effective action contains the finite part, the calculation of which
is extremely interesting, but is a more difficult and delicate
problem.

The divergent part of the effective action has been calculated
within the $\omega$ -regularization. If we use the other
regularization schemes, we can expect some extra terms in the
divergent part of the effective action. For example, the application
of the cut-off regularization results in the same two terms as in
(\ref{div}) only with a replacement of the term $\f{1}{\omega}$ with
$\sim logL^2$ where $L^2$ is the cut-off on the lower limit of the
proper-time integral (see the details e.g. in \cite{BK}). However,
within this regularization we will get the extra contribution to
divergent part of the effective action in the form \be S_{L^2}\sim
L^2\f{1}{4(4\pi)^3}\int d\zeta^{-4}duV^{++}Y^{++}.\ee This term is
generated from (\ref{integral}) when we take only the $Y^{++}$ in
the integrand, put $m^2=0$, $\omega = 0$ and cut the integral on the
lower limit by $L^2$. It is easy to see that this result is (up to a
coefficient) the Abelian (1,0) vector multiplet action (\ref{actZ}).

As a result we see that the classical actions (\ref{actZ}) and
(\ref{S}) of the Abelian theory are generated in quantum theory as
the one-loop counterterms. The Abelian higher derivative action
introduced in \cite{ISZ} is also generated as the one-loop
counterterm. We emphasize once more that a coupling to tensor
multiplet is stipulated by the superfield $\Phi$, at $\Phi = 0$ such
a coupling vanishes and (\ref{div}) gives us the divergent part of
the effective action for the hypermultiplet in a pure vector
multiplet background. It is worth pointing out that the superfield
calculation of the divergent part of the effective action is simple
enough in comparison with the component calculation and demonstrates
the power of superfield methods.

\section{Conclusion}

We have considered the superfield formulations of a class of six
dimensional supersymmetric models related to the low-energy dynamics
of $M5$ branes. These models possess the $\cN=(1,0)$ supersymmetry
and describe the hierarchy of interacting scalar, vector and tensor
fields and their superpartners \cite{SSWW}. Our main aim was to
construct the superfield actions for the above models. We have shown
that this aim has been achieved in the framework of a six
dimensional harmonic superspace. As a demonstration of the power of
the harmonic superspace approach we have considered the problem of
effective action in the hypermultiplet model coupled to the
background field of the vector/tensor system.

We have constructed the harmonic superfield Lagrangian formulation
of the free $\cN=(2,0)$ tensor multiplet in $\cN=(1,0)$ superspace.
The system of the (1,0) hypermultiplet and (1,0) tensor multiplet
was considered. The corresponding action is a sum of actions for the
corresponding (1,0) harmonic superfields. We have found the hidden
(1,0) supersymmetry transformation which mixes the hypermultiplet
and the tensor multiplet and have shown that this transformation
leaves the action invariant.

We have proposed the superfield Lagrangian formulation of the
non-Abelian tensor hierarchy in (1,0) harmonic superspace. The
superfield action has been formulated in terms of harmonic
superfields of vector and tensor multiplets. We reformulated the
constraints on the superstrengths \cite{IB} in terms of harmonic
superspace and by using these constraints we computed the component
action corresponding to the proposed superfield action. It was shown
that in an Abelian case this component action is analogous to the
action of the tensor hierarchy \cite{SSWW}.

To demonstrate a power of superfield methods we have considered a
problem of quantum effective action in Abelian hypermultiplet theory
coupled to background fields of vector/tensor system. Such an
effective action is generated by hypermultiplet loop and depends on
vector and tensor multiplet superfields. We have constructed the
second order differential operator with coefficients, depending on
background superfields, which acts on harmonic superfields and
defines a form of effective action. The superfield proper-time
technique for evaluating the effective action is developed. Such a
technique allows to compute the effective action in manifestly
supersymmetric and gauge invariant manner. We calculated the
divergent part of the effective action and showed that it has a
structure analogous to one of vector/tensor multiplet superfield
action and defines a renormalization of coupling constant. Also it
was shown that the actions of vector and tensor multiplet are
generated as the parts of divergences of the effective action.

There are various ways to generalize and apply the obtained results.
We will point out two of them. Firstly, in section 4 we constructed
the action of free (2,0) tensor multiplet in terms of (1,0)
hypermultiplet and tensor multiplets. The main element of this
construction was the existence of hidden (1,0) supersymmetry
transformations. We hope that such transformations can also be found
in the non-Abelian case which allows us to construct the non-Abelian
superfield action for the (2,0) tensor multiplet. Secondly, in
section 6 we began to study the effective action of the (1,0)
hypermultiplet coupled to the Abelian background field of the
vector/tensor system. We developed the superfield proper-time
technique for evaluating the effective action and calculated the
divergent part of the effective action. It would be extremely
interesting to find the finite part of this effective action since
it  be could a new $6D$ superconformal functional written in terms
of harmonic superspace. Also, it would be interesting to study the
effective action for the hypermultiplet in non-Abelian vector/tensor
background. We hope to consider the above problems in the
forthcoming papers.

\section*{Acknowledgments }

We are grateful to E.A. Ivanov, S.M. Kuzenko and I.B. Samsonov for
useful discussions. N.P. thanks I.A. Bandos for his help in the
early stages of the project. This work is supported in part by LRSS
grant, project No. 88.2014.2 and RFBR grant, project No.
12-02-00121-a.


\section
{Appendix A. Notations and conventions 
}

\refstepcounter{section}
\def\theequation{A.\arabic{equation}}
\setcounter{equation}{0}

In six dimensions the (1,0) and (0,1) Weyl spinors belong to the
fundamental representation  of $SU^\ast(4)\sim SO(1,5)$ group and to
the transpose representation, respectively. The $8\times 8$ Dirac
matrices $\Gamma^a$ (where $a=0,1,\ldots,5$ ) satisfy the Clifford
algebra
\be\label{DefG}\Gamma^a\Gamma^b+\Gamma^b\Gamma^a=2\eta^{ab}.\ee The
Dirac matrices for even dimensions can be chosen in the form 
$$\Gamma^a=\left(\begin{array}{cc}0 & (\g^a)_{\a\b}
\\(\tilde\g^a)^{\b\a} & 0 \\\end{array}\right),
$$
with $\a=1,\l, 4.$

Our notation and conventions follows to \cite{IB}. We use the metric
$\eta^{ab}=diag(+,-,-,-,-,-)$ as well as
$\ve_{abcdef}\ve^{a_1a_2a_3def}=-6\d^{a_1}_{[a}\d^{a_2}_b\d^{a_3}_{c]}.$
Everywhere the antisymmetrization with the weight 1 is used. We
chose the antisymmetric representation of the 6D Weyl $4\times 4$
$\g$-matrices $\g^a_{\a\b}=-\g^a_{\b\a}$ and
\be\tilde\g^{\a\b}_a=-\tilde\g^{\b\a}_a=\f12\ve^{\a\b\s\d}(\g_a)_{\s\d},\q
\g_{\a\b}^a=\f12\ve_{\a\b\s\d}(\tilde\g_a)^{\s\d},\ee where the
$SU^\ast(4)$ invariant  $\ve^{\a\b\s\d}$ is the totally
antisymmetric symbol ($\ve_{1234}=\ve^{1234}=1$). The matrices
$\g^a_{\a\b}$ obey the relation
\be\label{Defg}(\g^a\tilde\g^b+\g^b\tilde\g^a)_\a^{\ \
\b}=2\eta^{ab}\d_\a^{\ \ \b},\q
[\g_{ab},\g_c]=2\eta_{[bc}\g_{a]}.\ee The  six-dimensional
Pauli-type matrices $\{\g^a\}$ and $\{\tilde\g^a\}$ are two separate
bases of $4\times 4$ antisymmetric matrices so that \be
\g^a_{\a\b}\tilde\g_a^{\s\d}=-2\d_{[\a}^\s\d_{\b]}^\d, \q \q
\g^a_{\a\b}\g^a_{\s\d}=-2\ve_{\a\b\s\d},\q
\tilde\g_a^{\a\b}\tilde\g_a^{\s\d}=-2\ve^{\a\b\s\d}.\ee The
normalized antisymmetrized product of Pauli-type matrices \be
\g^{ab}=\f12(\g^a\tilde\g^b-\g^b\tilde\g^a),\q
(\tilde{\g_{ab}})^\a_{\ \ \b}=-(\g_{ab})_\b^{\ \ \a},\ee
$$\g_{abc}=\f{1}{3!}\g_{[a}\g_b\g_{c]}=\g_a\g_b\g_c-\g_a\eta_{bc}+\eta_{ac}\g_b-\eta_{ab}\g_c=\g^a\tilde{\g_{bc}}-\eta^{a[b}\g^{c]}=\g_{ab}\g_c+\eta_{c[a}\g_{b]},$$
$$ (\g^{abc})_{\a\b}=(\g^{abc})_{\b\a}=\f{1}{3!}\ve^{abcdef}(\g_{def})_{\a\b},\q (\tilde\g_{abc})^{\a\b}=
(\tilde\g_{abc})^{\b\a}=-\f{1}{3!}\ve_{abcdef}(\tilde\g_{def})^{\a\b},=\tilde\g_a\g_{bc}-\eta_{a[b}\tilde\g_{c]},$$
$$(\g^{abcdef})_\a^{\ \ \b}=-\ve^{abcdef}\d_\a^{\ \ b},
$$
$$(\g^{abcde})_{\a\b}=-\ve^{abcdef}(\g_f)_{\a\b},\q (\tilde\g^{abcde})^{\a\b}=\ve^{abcdef}(\tilde\g_f)^{\a\b},$$
$$(\g^{abcd})_{\a\b}=\f12\ve^{abcdef}(\g_{ef})_{\a\b},\q (\tilde\g^{abcd})^{\a\b}=-\f12\ve^{abcdef}(\tilde\g_{ef})^{\a\b},$$
form the basis of general $4\times 4$ matrices with the completeness
relation \be(\g^{ab})_\a^{\ \ \b}(\g_{ab})_\s^{\ \
\d}=2\d_\a^\b\d_\s^\d-8\d_\a^\d\d_\s^\b,\q\q
\g^{abc}_{\a\b}\tilde\g^{\s\d}_{abc}=-24\d_\a^{(\s}\d_\b^{\d)}, \q
\g^{abc}_{\a\b} \g_{\s\d}^{abc}=0,\ee
$$\g^a_{\a\b}(\g_{ab})_\d^\rho=2\d_\a^\rho\g^b_{\b\d}+2\d_\b^\rho\g^b_{\d\a}+\d_\d^\rho\g^b_{\a\b},\q \tilde\g_a^{\a\b}(\g_{ab})_\d^\rho=2\d_\d^\b\tilde\g_b^{\a\rho}+
2\d^\a_\d\tilde\g_b^{\rho\b}-\d_\d^\rho\tilde\g_b^{\a\b},$$
$$\g^a_{\a\b}(\g_{abc})_{\g\d}=2\ve_{\a\b\g\s}(\g_{bc})_\d^{\ \ \s}-\g^{[b}_{\a\b}\g^{c]}_{\g\d},\q \g^a_{\a\b}(\tilde\g_{abc})^{\g\d}=-2\d^\g_{[\a}(\g_{bc})^\d_{\b]}
-\g_{[b \a\b}\tilde\g_{c]}^{\g\d},$$
$$(\tilde\g_{abc})^{\a\b}(\g_{ab})_\b^\rho=20\tilde\g_c^{\rho\a}, \q \g^a_{\a\d}(\tilde\g_{abc})^{\d\b}=4(\g_{bc})_\a^\b,$$
$$(\g_{abc})_{\a\b}(\g^{ab})_\g^\d=4\d_\a^\d\g^c_{\b\g}+4\d_\b^\d\g^c_{\a\g}.$$
The trace relations are \be \tr(\g^a\tilde\g^b)=4\eta^{ab},\q
\tr\g_{ab}\g^{cd}=-4\d^c_{[a}\d^d_{b]},\q
\tr(\g_{abc}\tilde\g^{def})=-4\ve^{abc}_{\ \ \
def}-4\d^a_{[d}\d^b_e\d^c_{f]},\ee
$$\tr(\g^a\tilde\g^b\g^c\tilde\g^d)=4(\eta^{ab}\eta^{cd}-\eta^{ac}\eta^{bd}+\eta^{ad}\eta^{bc}),\q \tr(\g_e\tilde\g_{abc})=0,$$
$$\tilde\g_a\g^b\tilde\g_a=-4\tilde\g^b,\q \g_e\g_{ab}\g_e=2\g_{ab},\q \g_e\g_{abc}\g_e=0, \q \g_e\g^{def}\g_{ab}\g_e=-2\g_{[a}\g^{def}\g_{b]}.$$
A Minkowski six-vector can be written as the bi-spinor: \be
x_{\a\b}=\g^a_{\a\b}x_a, \q x^{\a\b}=\tilde\g_a^{\a\b}x^a,\q
x^2=\f14x^{\a\b}x_{\b\a}, \q x_a= \f14\tr(\tilde\g_a x)\equiv
\f14\tilde\g_a^{\a\b}x_{\b\a},\ee
$$\p_{\a\b}=\g^a_{\a\b}\p_a,\q \p_{\a\b}x^{\g\d}=-2\d^\g_{[\a}\d^\d_{\b]},\q \p_{\a\b}x_{\g\d}=-2\ve_{\a\b\g\d}.$$
The supersymmetric covariant derivatives in the central basis of the
(1,0) D6 harmonic superspace have the form \be
D^+_\a=\f{\p}{\p\t^{-\a}}-i\p_{\a\b}\t^{+\b },\q D^-_\a=-\f{\p}{\p\t^{+\a}}-i\p_{\a\b}\t^{-\b },  \q \{D^+_\a,
D^-_\b\}=2i\p_{\a\b}
.\ee The definition of the vector superfield strength looks like
\be[\cD_a,\cD_b]=F_{ab},\q\q \{\cD_\a^{(i},\cD_\b^{j)}\}=0,\ee
$$F_{ab}=-\f18(\g_{ab})_\a^{\ \b}F_\b^{\ \a }, \q F_\a^{\ \b}=(\g_{ab})_\a^{\ \b}F^{ab}.$$
The field strength of the 2-form potential can be decomposed as
follows \be H^\pm_{abc}=\f12(H_{abc}\pm \ast H_{abc}),\q \ast
H_{abc}=\f16\ve_{abcdef}H^{def},\ee where in the spinor
representation the (anti-)self-dual parts of a 3-form $H$ satisfy
the relations
$$H^{(-)}_{\a\b}=H_{abc}(\g^{abc})_{\a\b}, \q H^{(+)\a\b}=H^{abc}(\g_{abc})^{\a\b}.$$
We also used the following notations and conventions
\be(D^\pm)^4=-\f{1}{4!}\ve^{\a\b\rho\g}D^\pm_\a D^\pm_\b D^\pm_\rho
D^\pm_\g , \q (D^+)^{3\a}=-\f16\ve^{\a\b\g\d}D^+_\b D^+_\g D^+_\d
,\ee
$$D^\pm_\a D^\pm_\b D^\pm_\rho =\ve_{\a\b\rho\g}(D^\pm)^{3\g}, \q D^\pm_\a
D^\pm_\b D^\pm_\g D^\pm_\d=-\ve_{\a\b\g\d}(D^\pm)^{4},\q
D^\pm_\a(D^\pm)^{3\b}=\d_\a^\b(D^\pm)^4,
$$
and also
\be\t^{\pm\a}\t^{\pm\b}\t^{\pm\g}=-\ve^{\a\b\g\d}(\t^\pm)_\d^3, \q (\t^\pm)^3_\a=\f16\ve_{\a\b\g\d}\t^{\pm\b}\t^{\pm\g}\t^{\pm\d},\ee
$$ \t^{\pm\a}\t^{\pm\b}\t^{\pm\g}\t^{\pm\d}=-\ve^{\a\b\g\d}(\t^\pm)^4, \q
(\t^\pm)^4=-\f{1}{4!}\ve_{\a\b\g\d}\t^{\pm\a} \t^{\pm\b} \t^{\pm\g}
\t^{\pm\d},\q \t^{\pm\a}(\t^\pm)^3_\b=-\d^\a_\b(\t^\pm)^4,$$
$$(D^+)^{3\a}(\t^-)^3_\b=\d_\b^\a, \q (D^+)^4(\t^-)^4=1,\q (D^+)^{3\a}(\t^-)^4=\t^{-\a}.$$


\section
{Appendix B. Component expansion of the action (\ref{S})
}

\refstepcounter{section}
\def\theequation{B.\arabic{equation}}
\setcounter{equation}{0}

We will now consider  the main steps to derive the component
decomposition of the action $S\sim \int d\zeta^{(-4)}du{\cL}^{(+4)}$
in the Abelian case. Here ${\cL}^{(+4)}$ is given by (\ref{S}). To
do that, we should integrate over harmonics and over all
anticommuting coordinates.

Let us begin with the integration rule over anticommuting
coordinates \be \int d\zeta^{(-4)}du= \int d^6x du
(-\f{1}{4!})\ve^{\a\b\g\d}\cD^-_\a \cD^-_\b \cD^-_\g \cD^-_\d.\ee
First, we act by spinor derivatives and kill all the theta's and
then integrate over harmonics $u^\pm_i.$ It is obvious that
$\cD^-_\a$ does not act on $\t^{-\a}$, therefore the dependence on
them in (\ref{S}) can be omitted from the very beginning. Using the
rules $[\cD^{++},\cD^-_\a]=\cD^+_\a$ in the expression
 $\cD^-_\a \cD^-_\b \cD^-_\g \cD^-_\d{\cal L}^{(+4)}$
we get a number of terms which are conveniently grouped into
\be\label{derPhi}-\Phi(D^+_\a D^-_\b D^-_\g D^-_\d Y^{++}+D^-_\a
D^+_\b D^-_\g D^-_\d Y^{++}+D^-_\a D^-_\b D^+_\g D^-_\d Y^{++})\ee
\be\label{derPsi}+4D^-_\a\Phi D^{++}D^-_\b D^-_\g D^-_\d
Y^{++}-4D^-_\a\Phi(D^+_\b D^-_\g D^-_\d Y^{++}+D^-_\b D^+_\g D^-_\d
Y^{++})\ee
$$+D^+_\a\Phi D^-_\b D^-_\g D^-_\d Y^{++}-D_\r^+\Phi (D^+_\a D^-_\b D^-_\g D^-_\d+D^-_\a D^+_\b D^-_\g D^-_\d+ D^-_\a D^-_\b D^+_\g D^-_\d)W^{+\r}$$
\be\label{calH} \{+6D^-_\a D^-_\b \Phi D^{++}D^-_\g D^-_\d Y^{++}- 6D^-_\a
D^-_\b \Phi D^+_\g D^-_\d Y^{++}\}+4D^-_\a D^+_\b\Phi D^-_\g
D^-_\d  Y^{++} \ee
$$+4 D^-_\a D^+_{\r}\Phi(D^+_\b D^-_\g D^-_\d +D^-_\b D^+_\g D^-_\d )W^{+\r}
-4D^-_\a  D_\r^+\Phi D^{++}D^-_\b D^-_\g D^-_\d W^{+\r} $$
$$\{+6D^-_\a D^-_\b D^+_\g\Phi  D^-_\d  Y^{++}+6D^-_\a D^-_\b D^+_\r\Phi D^{++} D^-_\g D^-_\d W^{+\r} -6D^-_\a D^-_\b D_\r^+\Phi D^+_\g D^-_\d W^{+\r}\}$$
\be\label{zero}-4D^-_\a D^-_\b D^-_\g D_\r^+\Phi D^{++} D^-_\d
W^{+\r}+4D^-_\a D^-_\b D^-_\g D^+_\d\Phi    Y^{++}\ee
$$+\Phi D^{++}D^-_\a D^-_\b D^-_\g D^-_\d Y^{++}+D^-_\a D^-_\b D^-_\g D^-_\d\Phi D^{++} Y^{++}$$
$$+D_\r^+\Phi D^{++}D^-_\a D^-_\b D^-_\g D^-_\d W^{+\r} + D^-_\a D^-_\b D^-_\g D^-_\d D_\r^+\Phi D^{++} W^{+\r}
+3D^-_\a D^-_\b D^-_\g \Phi D^{++}D^-_\d Y^{++}~.$$ Further we will
investigate each of these expressions separately.

First, note that the set of terms (\ref{zero}) obviously vanishes if
we recall the properties of harmonic superfields. In order to
transform the expression (\ref{derPhi}) to the component form we
should commute the spinor derivative $D^+_\a$ to the right, use the
fact that $D^+_\a Y^{++}=0$ and take into account that
$[\cD_\g^-,\cD_{\a\b}]=-2i\ve_{\a\b\g\d}W^{-\d}.$  As a result one
gets
$$\f{1}{4!}\ve^{\a\b\g\d}\Phi(12iD_{\a\b}D^-_\g D^-_\d +32\ve_{\a\b\g\r}W^{-\r} D^-_\d+12\ve_{\b\g\d\r}(D^-_\a W^{-\r}) )Y^{++}~.$$
The equations (\ref{DY}), (\ref{DP}) of subsection 5.2  allow us to
rewrite this equation as follows:
$$\Phi(D^{\a\b}D_{\a\b}\Phi-16iW^{-\a}D_{\a\b}W^{+\b}-12Y^{--}Y^{++}+32iW^{-\a}\Psi^+_\a-\f12{\cF}_{\a}^{\ \b}{\cF}_{\b}^{\ \a}
-2D^{\a\b}D_{\a\b}Y^{+-}).
$$
The last term of the above relation vanishes after integration over
the harmonic variables $\ve_{ij}Y^{ij}\equiv 0$. The obvious
transformations of the other terms under the integral $\int du u^+_i
u^-_j=\f12\ve_{ij}$, $\int du u^+_iu^+_ju^-_k
u^-_l=\f16\ve_{(ik}\ve_{j)l}$ give \be\int d^6x \{4D^a\Phi
D_a\Phi+\Phi(4{\cF}^{ab}{\cF}_{ab}-4Y^{ij}Y_{ij}-8i W^{\a}_i
D_{\a\b}W^{i\b}+16iW^{\a}_i\Psi^i_\a)\}.\ee

The other  terms in the expression (\ref{derPsi}) are considered
analogously. As a result, one gets
$$ -24i\Psi^+_\a W^{+\a}Y^{--}+16i\Psi^+_\a
W^{-\a}(Y^{+-}+\f12\Phi)+16i\Psi^+_\a D^{\a\b}\Psi^-_\b-16i\Psi^-_\a
D^{\a\b}\Psi^+_\b$$
$$ -8i\Psi^+_\a D^{\a\b}D_{\b\g}W^{-\g}-4i\Psi^+_\r
D^{\a\b}D_{\a\b} W^{-\r}+8i\Psi^-_\a D^{\a\b}D_{\b\g}W^{+\g}~.$$
Integrating over harmonics leads to  \be+4i\Psi^i_\a
W_i^{\a}\Phi+16i\Psi^i_\a D^{\a\b}\Psi_{i\b}+4i\Psi_\a^i
W^\b_i{\cF}_\b^{\ \ \a}-\f{32}{3}i\Psi^i_\a W^{j\a}Y_{ij}~.\ee

The expression (\ref{calH}) has a complicated structure. However,
bear in mind that we have the properties $W^{\a i}W^\b_i=W^{\b
i}W^\a_i$ in the Abelian case. This allows us to make cancellations
of the potentially admissible terms
$\ve_{\a\b\g\d}W^{-\a}W^{+\b}W^{-\g}W^{+\d}$, $D_{\a\b}\Phi
W^{-\a}W^{+\b}$. As a result, we have
$$-\f{1}{12}{\cH^{(-)}}_{\a\b}D^{(\a\d}{\cF}_\d^{\ \ \b)}-
\f{8i}{3}{\cH}^{(-)}_{\a\b}W^{-\a}W^{+\b}+D^{\a\b}\Phi
D_{\b\r}{\cF}_\a^{\ \r}- D_{\a\b}\Phi D^{\a\d}{\cF}_\d^{\ \b},$$
where the last two terms disappear. Finally, after using the
identity (\ref{H}) this expression takes the form \be \int d^6x \{
-\f{1}{18}{\cH^{(-)}}_{\a\b}{\cH^{(+)}}^{\a\b}-\f{4i}{3}{\cH}^{(-)}_{\a\b}
W^{\a}_i W^{j\b}\}~. \ee

Thus we see that all the functional structures of the component
action (\ref{component}) are obtained from the superfield action
(\ref{S}).

\bigskip

\end{document}